\documentclass[final,11pt]{elsarticle}

\usepackage{amssymb,amsmath,amssymb}

\usepackage{mathrsfs}
\usepackage{float}
\usepackage{CJK}
\usepackage{indentfirst}
\usepackage{amsmath}
\usepackage{cases}

\usepackage{amsthm}
\usepackage{graphics}
\usepackage{caption2}
\usepackage{psfrag}
\usepackage{showlabels}
\usepackage{color}
\usepackage{amsmath}
\usepackage{bm}
\usepackage{mathtools}
\usepackage{tensor}
\usepackage{subfigure}
\usepackage[normalem]{ulem}
\usepackage[linesnumbered,boxed,ruled,commentsnumbered]{algorithm2e}
\usepackage{graphicx}
\usepackage{epstopdf}
\usepackage{array}
\usepackage{makecell}
\usepackage[english]{babel}
\usepackage{booktabs}
\usepackage{threeparttable}

\usepackage{amsfonts}
\usepackage{amssymb}
\usepackage{graphicx}
\usepackage{multirow}
\usepackage{exscale}
\usepackage{relsize}
\usepackage[normalem]{ulem}
\usepackage{color}
\usepackage{fullpage}
\usepackage{lineno}
\allowdisplaybreaks[4]

\newcommand{\ds}{\displaystyle}

\newcommand{\x}{\mathbf{x}}

\newcommand{\be}{\begin{equation}}
\newcommand{\ee}{\end{equation}}
\newcommand{\ba}{\begin{array}}
\newcommand{\ea}{\end{array}}

\newtheorem{remark}{Remark}
\allowdisplaybreaks[4]
\usepackage{lineno}
\UseRawInputEncoding

\journal{Computer Methods in Applied Mechanics and Engineering}

\begin{document}
\begin{frontmatter}

\title{An asymptotically compatible  bond-based peridynamics with Gaussian kernel}
\cortext[cor1]{Corresponding author.}
\author[ad1]{Hao Tian}
\ead{haot@ouc.edu.cn}
\author[ad1]{Jinlong Shao}
\ead{shaojinlong@stu.ouc.edu.cn}
\address[ad1]{School of Mathematical Sciences, Ocean University of China, Qingdao, Shandong 266100, China}
\author[ad1]{Chenguang Liu\corref{cor1}}
\ead{liuchenguang@stu.ouc.edu.cn}

\begin{abstract}
In this paper, we introduce a novel bond-based peridynamic model that utilizes a Gaussian kernel function. Previous peridynamic models, when directly discretized, have exhibited a lack of asymptotic compatibility with their corresponding local elastic solutions. Additionally, these models have faced challenges in accurately computing the volume of intersecting regions between the horizon of the material point and its neighboring cells. These difficulties in numerical simulations within peridynamics have spurred numerous efforts to develop corrective numerical methods. While such corrective methods have addressed certain issues, they remain complex to formulate and computationally intensive.
Instead of addressing these challenges through modified numerical discretization, this paper presents a novel approach: bond-based peridynamics with a Gaussian kernel (BBPD-GK). This model replaces the bounded kernel function commonly used in existing peridynamic models with an unbounded Gaussian kernel function. Through direct meshfree discretization, we demonstrate that the BBPD-GK solution, without the need for compatibility correction or volume correction, aligns with the corresponding local elastic solution. Furthermore, we derive the relevant material parameters based on energy equivalence to an elastic continuum model. Building upon this foundation, we propose a damage model and a linearized version of BBPD-GK. We validate our proposed model through simulations of 2D smooth problems, demonstrating its convergence and accuracy. Finally, we assess the applicability of our approach to realistic scenarios by predicting displacements caused by a 2D plate with pre-existing cracks and replicating high-velocity impact results from the Kalthoff-Winkler experiments.
\end{abstract}


\end{frontmatter}

\section{Introduction}

Classical continuum mechanics are often expressed as a system of partial differential equations, which is challenging to describe models with discontinuities. As a result, peridynamics (PD), as proposed by Silling \cite{SA}, reformulate the motion equations of classical mechanics based on nonlocal theory and can solve the deformation problems of materials with a discontinuity. The initially developed model was the bond-based peridynamics model (BB-PD), which could only handle brittle materials with a fixed Poisson's ratio. Subsequently, the proposal of ordinary state-based peridynamics models (OSB-PD) \cite{SA2} enabled PD to handle isotropic materials with arbitrary Poisson's ratios, and nonordinary stated-based model (NOSB-PD)  was established to handle anisotropic materials. 
Over the past decades, the effectiveness of PD has attracted extensive research and has been frequently used in many practical problems, such as composite material deformation \cite{Madenci2,Han,Liu1,Liu2}, corrosion \cite{Zhang3,SZF,ZSJ}, ocean engineering\cite{Ocean1,Ocean2,Ocean3}, and crack simulation\cite{Madenci1,Zhang1,Zhang2,Guo1}. Significant research has been dedicated to developing numerical methods, including meshfree, finite difference, finite element, and collocation, to solve PD models \cite{TCP, XM, SilAsk, JY, JMY, SY}.

Although peridynamics (PD) reformulates the equations of motion in classical mechanics, it can still require to be compatiable to classical mechanics. The initial PD model, introduced by Silling \cite{SE}, is a nonlocal extension of the classical continuum mechanics model, with relevant material parameters derived from energy equivalence to classical continuum mechanics. In \cite{DU2011}, Du indicated that the solution of PD is compatible with that of the Navier equation when the nonlocal effect diminishes, and evidence of compatibility between the solutions of linear state-based PD and the classical Navier-Lamé system is provided in \cite{Du2014}. However, according to \cite{DU2013}, existing numerical methods fail to provide compatibility between the solutions of PD and classical mechanics, even with an increase in the number of grids. To solve this problem, \cite{DU2013} and \cite{Ti2014} developed corrective numerical methods for the one-dimensional bond-based PD model (BB-PD), enabling the solution to be compatible with the expected local solution. Subsequently, \cite{Yang2018} proposed a corrective compatible discretization for multidimensional PD diffusion models. Nevertheless, these correction methods are difficult to design and lack approaches for multidimensional BB-PD. By using local optimization problems to seek quadrature weights associated with a local neighborhood of particles, a compatible meshfree discretization method for multidimensional BB-PD was introduced in \cite{AC2}. In \cite{AC3}, quadrature weights were calculated using the generalized moving least squares technique, and a meshfree integration scheme was proposed to obtain a PD solution compatible with the classical mechanics model. However, these methods entail a significant computational burden.

The accurate evaluation of intersections for the direct discretization method also poses a significant challenge hindering the application of peridynamics (PD). Theoretically, the calculation of PD forces should use information from all material points within the nonlocal neighborhood. Material points located near the boundary of the neighborhood partially belong to nonlocal neighborhoods, making it difficult to calculate their volume. The initial method to address this issue, proposed in \cite{MF}, considers the volume of material points in the boundary region as the complete volume of material points, which is not very accurate. Subsequently, a series of numerical methods were developed to correct this influence, such as the QWJ algorithm \cite{vol3}, the PA-AC algorithm \cite{vol4}, and the VCA algorithm \cite{vol5}. However, these methods heavily depend on the type of discretization used. Additionally, introducing volume correction also incurs extra computational costs. 

The primary contribution of this study lies in proposing an asymptotically compatible peridynamic model by redefining the traditional bond-based peridynamic model. The proposed model constructs the kernel function using the Gaussian function and can achieve asymptotic compatibility convergence without the need for specialized numerical techniques. The horizon for the novel model is an unbounded region, and to lessen the computational burden in practical applications, we chose a truncation to replace the influence region. Consequently, our model maintains accuracy without the need for volume correction methods.  By energy equivalence to a classical linear elastic continuum mechanics model, the relevant material in the nonlinear model is introduced, and the damage model is constructed. The linear model is proposed by introducing the micromodulus function and verifying the convergence by theoretical analysis. Meshfree methods were chosen as representatives of spatial discretization techniques, and we validated the accuracy and convergence of the proposed model across both smooth and discontinuous problems. The innovation point of this work is as follows:
\begin{itemize}
\item To the best of our knowledge, this is the first instance of presenting an asymptotically compatible bond-based peridynamic model, whose straightforward meshfree discretization aligns with the corresponding local elastic solution without requiring any numerical correction.

\item By using unbounded Gaussian kernel functions and applying appropriate truncation, we obviate the requirement for volume correction algorithms when solving peridynamic problems.

\item The corresponding linearized model and damage model have been proposed, and their effectiveness is examined through a variety of numerical experiments.
\end{itemize}

The structure of the remaining sections of this paper is as follows. Section \ref{section2} provides an overview of the traditional bond-based peridynamic model. In Section \ref{section3}, we introduce the formulation of the nonlinear novel bond-based peridynamic model and conduct crack identification analysis through constitutive equations.  Section \ref{section4} presents the linear novel bond-based peridynamic model and analyzes its asymptotically compatible convergence through theoretical analysis. The 3D BBPD-GK model are introduced in Section \ref{section5}. In Section \ref{section6}, we describe the spatial discretization method for the novel model. The accuracy and convergence of the proposed model are validated using numerical examples in Section \ref{section7} and \ref{section8}.

\section{Peridynamic model of a continuum}
\label{section2}
As a nonlocal model, the primary PD encompasses three types: bond-based peridynamics (BB-PD), ordinary state-based peridynamics (OSB-PD), and non-ordinary state-based peridynamics (NOSB-PD). In this paper, our emphasis primarily revolves around BB-PD. \begin{equation}\label{pd:e1}
\vspace{0.01in}
\rho \ddot{\mathbf{u}}(\mathbf{x}, t)= \ds \int_{\mathcal{R}} \mathbf{f}(\bm{\eta},\bm{\xi}) dV_{\mathbf{\mathbf{x}^{\prime}}}+\mathbf{b}(\mathbf{x},t), 
\end{equation}
where $\rho$ denotes the mass density, $\mathbf{b}(\mathbf{x},t)$ represents the body force density field, $\bm{\xi}=\mathbf{\mathbf{x}^{\prime}}-\mathbf{\mathbf{x}}$,  $\bm{\eta}=\mathbf{u}(\mathbf{\mathbf{x}^{\prime}},t)-\mathbf{u}(\mathbf{\mathbf{x}},t)$, with   $\mathbf{u}(\mathbf{x}^{\prime},t)$ and $\mathbf{u}(\mathbf{x},t)$ represent displacemnts of the material points of location $\mathbf{\mathbf{x}^{\prime}}$ and $\mathbf{\mathbf{x}}$, respectively. $\mathbf{f}(\bm{\eta},\bm{\xi})$ denotes a pairwise force function used to compute the force vector and satisfies $\mathbf{f}(\bm{\eta},\bm{\xi})=-\mathbf{f}(-\bm{\eta},-\bm{\xi})$, which can be expressed as the follows:
\begin{equation}
	\mathbf{f}(\bm{\eta},\bm{\xi})=\begin{cases}
	\dfrac{\boldsymbol{\xi}+\boldsymbol{\eta}}{|\boldsymbol{\xi}+\boldsymbol{\eta}|}c(\bm{\xi}) s \mu,& |\bm{\xi}|\leq \delta,\vspace{0.5em}\\
	0,&otherwise,
	\end{cases}
\end{equation}
  $c$  is a function representing the micromodulus of  the bond stiffness,  which conclude follows for plane stress cases: 
 \begin{equation}\label{bc}
     \begin{aligned}
  \textbf{constant:}\quad &c=c(\bm{\xi})=\frac{9E}{\pi\delta^{3}h},\\
 \textbf{conical:}\quad &c(\bm{\xi})= \frac{27E}{\pi\delta^{3}h}\left(1-\frac{|\bm{\xi}|}{\delta}\right), 
\end{aligned}
 \end{equation}
$s$ is the PD bond stretch, $\mu$ is  is a scalar representing the current status of the PD bond, which can be expressed as 
\begin{equation}\label{fra:s1}
 s=\frac{|\boldsymbol{\xi}+\boldsymbol{\eta}|-|\boldsymbol{\xi}|}{|\boldsymbol{\xi}|}, \quad   \mu= \begin{cases} $1$, &\text{ $s<s_0$}, \\ $0$, & \text { $s \ge s_0$},\end{cases} \quad s_0= \displaystyle{\begin{cases}\sqrt{\frac{4 \pi G_0}{9 E \delta}}, & \text {2D,} \\ \sqrt{\frac{5 \pi G_0}{12 E \delta}}, & \text {3D,} \end{cases}}
\end{equation}
where $h$ is the thick, $E$ is the elasticity modulus, $G_0$ being the energy release rate.

Typically, when  displacement is sufficiently small, nonlinear model can be equal to linear bond-based peridynamics, which can be expressed as follows
\begin{equation}
		\mathbf{f}(\bm{\eta},\bm{\xi})=\begin{cases}c(\bm{\xi})\mu\dfrac{\boldsymbol{\xi}\otimes\boldsymbol{\xi}}{|\boldsymbol{\xi}|^3}\bm{\eta},& |\bm{\xi}|\leq \delta,\vspace{0.5em}\\
		0,&otherwise,
		\end{cases}
\end{equation}
\section{Two-dimension Gaussian-kernel bond-based peridynamics}
\label{section3}
   In this section, a novel bond-based peridynamics with Gaussian kernel (BBPD-GK) is introduced, and a nonlinear model is given. First,  a novel nonlocal operator is introduced to construct the novel BB-PD
\begin{equation}
\mathcal{L}_{\sigma}\mathbf{u}(\mathbf{x},t)=\int_{\mathcal{R}}\mathbf{g}(\bm{\eta},\bm{\xi}) dV_{\mathbf{\mathbf{x}^{\prime}}}+\mathbf{b}(\mathbf{x},t).
\end{equation}
here  $\mathbf{g}(\bm{\eta},\bm{\xi})$ is the force vector and satisfy  
\begin{equation}
    \mathbf{g}(\bm{\eta},\bm{\xi})=-\mathbf{g}(-\bm{\eta},-\bm{\xi}),\quad \mathbf{g}(\bm{\eta},\bm{\xi})=\dfrac{\boldsymbol{\xi}+\boldsymbol{\eta}}{|\boldsymbol{\xi}+\boldsymbol{\eta}|}g(\bm{\eta},\bm{\xi}).
\end{equation}
For two-dimension, $g(\bm{\eta},\bm{\xi})$ is the scalar-valued function can be expressed as
\begin{equation}\label{pde3}
    g(\bm{\eta},\bm{\xi})=\beta s c(\bm{\xi}),
\end{equation}
and  $c(\bm{\xi})$ is chosen as
\begin{equation}
    c(\bm{\xi})=G(\bm{\xi})|\bm{\xi}|^3,
\end{equation}
 where $\beta$ is a constant,  $G(\bm{\xi})$ is a kernel function which can be expressed as
 \begin{equation}
 G(\bm{\xi})=\dfrac{1}{2\pi\sigma^2}exp\left({-\dfrac{|\bm{\xi}|}{2\sigma^2}}\right),
 \end{equation}
 with $\sigma$ is a constant represented variance. 

\begin{figure}[ht]
	\centering    	
	\subfigure[]
	{
		\begin{minipage}{7cm}
			\centering          
			\includegraphics[scale=0.4]{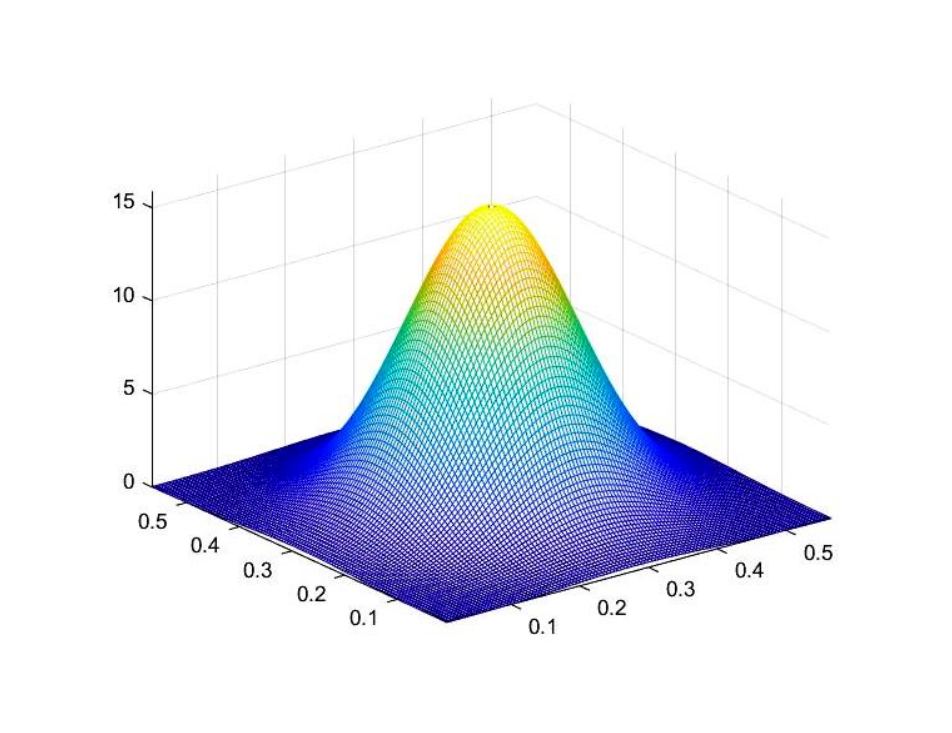}  
		\end{minipage}
	}	
	\subfigure[]
	{
		\begin{minipage}{7cm}
			\centering      
			\includegraphics[scale=0.4]{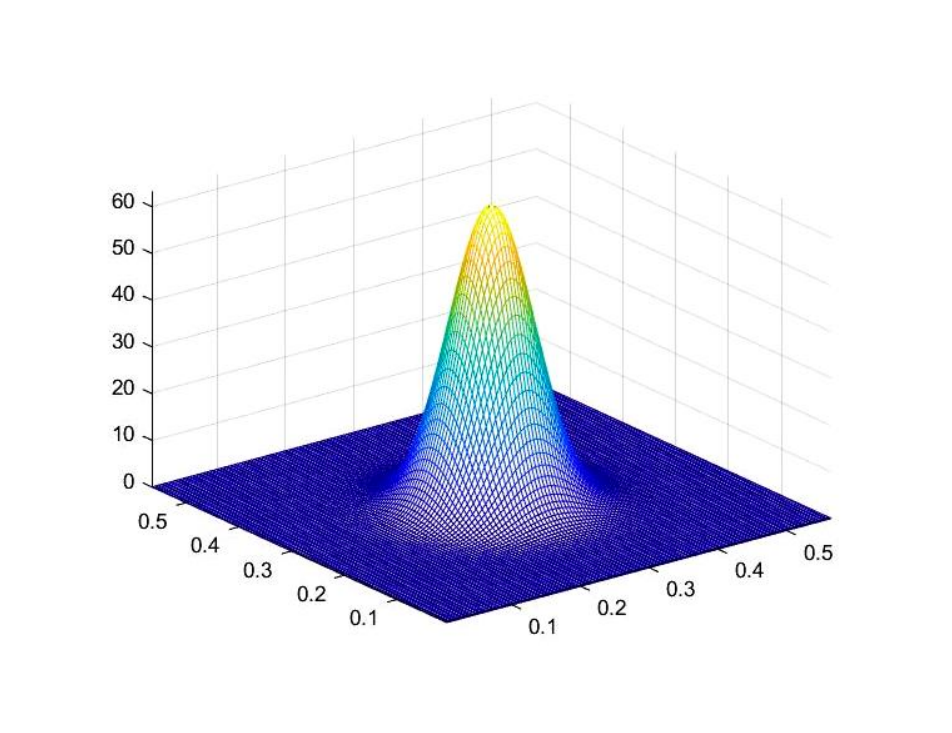}   
		\end{minipage}
	}	
	\caption{ Plots of the Gaussian function $G(\bm{\xi})$ (in two dimensions) with $\sigma= 1/10$ (left) and $\sigma= 1/20$ (right), respectively. 
} 
	\label{Fig:sigma}
\end{figure}

Here, $G(\bm{\xi})$ is a two-dimensional multivariate Gaussian distribution with variance $\sigma$, which represent the strength of relation between $\mathbf{x}^{\prime}$ with $\mathbf{x}$.  It is easy to find that the shape of $G(\bm{\xi})$is decided by the variance $\sigma$, and the smaller $\sigma$, the more elongated and concentrated around the center, as shown in Fig. \ref{Fig:sigma}.

It should be noted that an essential difference between BBPD and BBPD-GK lies in the choice of kernel function.
In traditional BB-PD, the kernel function is bounded, and the material points interact within
a disk with a radius of $\delta$. However the kernel function in this model has an unbounded support region, the shape of which is totally determined by the variance $\sigma$.

As shown in Fig. \ref{Fig:sigma}, one should note is that although the influence region of $G(\bm{\xi})$ for a point $\mathbf{x}$ is unbounded, but  as the distance between $\mathbf{x}$ and other material points decreases, the interaction between material points also weakens to 0. This means that in practical calculations, a bounded truncated region can be used to replace an unbounded region.
\subsection{Constitutive modeling}
 In this section, we will obtain the parameter $\beta$ for the 2D model by constitutive modeling. Similar to the BB-PD, the function $\mathbf{g}(\bm{\eta},\bm{\xi})$ also satisfy
 \begin{equation}
         \mathbf{g}(\bm{\eta},\bm{\xi})=\dfrac{\partial \omega(\bm{\xi},\bm{\eta})}{\partial \bm{\eta}},
 \end{equation}
where $\omega(\bm{\xi},\bm{\eta})$ can be obtained by
\begin{equation}
	W=\frac{1}{2} \int_{\mathcal{R}} w(\boldsymbol{\eta}, \boldsymbol{\xi}) \mathrm{d} V_{\xi}.
\end{equation}

Consider an isotropic homogeneous body, where $s$ is constant for all $\xi$, and $\bm{\eta}=s \bm{\xi}$. Let $\xi=|\bm{\xi}|$ and $\eta=|\boldsymbol{\eta}|$, we can obtain $\eta=s \xi$. Since $g=\beta s G(\bm{\xi})\xi^3=\beta \eta G(\bm{\xi})\xi^2$, it follows that $w=\beta \eta^2\xi^2 G(\bm{\xi})/ 2 =\beta s^2 \xi^4 G(\bm{\xi}) / 2 $. Then for 2D BBPD-GK model, we have
  \begin{equation}
  	  W=\frac{1}{2}\int_\mathcal{R} w dV_{\mathbf{\mathbf{x}^{\prime}}}=
  	\frac{1}{2} \int_0^\infty \dfrac{\beta s^2 \xi^4 G(\bm{\xi})}{2} 2 \pi \xi d\xi=2\beta s^2\sigma^4 ,
  \end{equation}
According to \cite{MF}, we have the following equality
    \begin{equation}
  	W=2\beta s^2\sigma^4=\dfrac{3Es^2}{2},
  \end{equation} 
  thus we have
   \begin{equation}
 	\beta=\dfrac{3E}{4\sigma^{4}},
 \end{equation}
 which is called the spring constant in the PMB material.
 \subsection{Notion of failure}
 In this section, we will introduce the failure criteria of BBPD-GK. Similar to BB-PD, a history-dependent scalar-valued function similar to \eqref{fra:s1} is introduced, and the force density vector $\mathbf{g}$ in BBPD-GK can be rewritten as
 \begin{equation}\label{gnolin}
 	\mathbf{g}(\boldsymbol{\eta}, \boldsymbol{\xi})=\mu\dfrac{\boldsymbol{\xi}+\boldsymbol{\eta}}{|\boldsymbol{\xi}+\boldsymbol{\eta}|} g(\boldsymbol{\eta}, \boldsymbol{\xi}),\quad \mu= \begin{cases} 1, &\text{ $s\hspace{0.3em}<\hspace{0.3em} s_c$}, \\ 0, & \text { $s \ge s_c$},\end{cases},
 \end{equation}
 Here, $s_c$ is the critical stretch for bond failure and usually take a constant. Then, the notion of local damage can be  defined as
 \begin{equation}
 	\varphi(\mathbf{x}, t)=1-\dfrac{\int_{\mathcal{R}} \mu(\mathbf{x}, t, \boldsymbol{\xi}) \mathrm{d} V_{\xi}}{\int_{\mathcal{R}} \mathrm{d} V_{\xi}}
 \end{equation}
 where $\mathbf{x}$ is now included as an argument of $\mu$ as a reminder that it is a function of position in the body. When $\varphi=0$, it means that this is still the original material, and when $\varphi=1$, it represents complete disconnection of a point from all of the points with which it initially interacted.

\begin{figure}[ht]
		\centering            
		\includegraphics[scale=0.5]{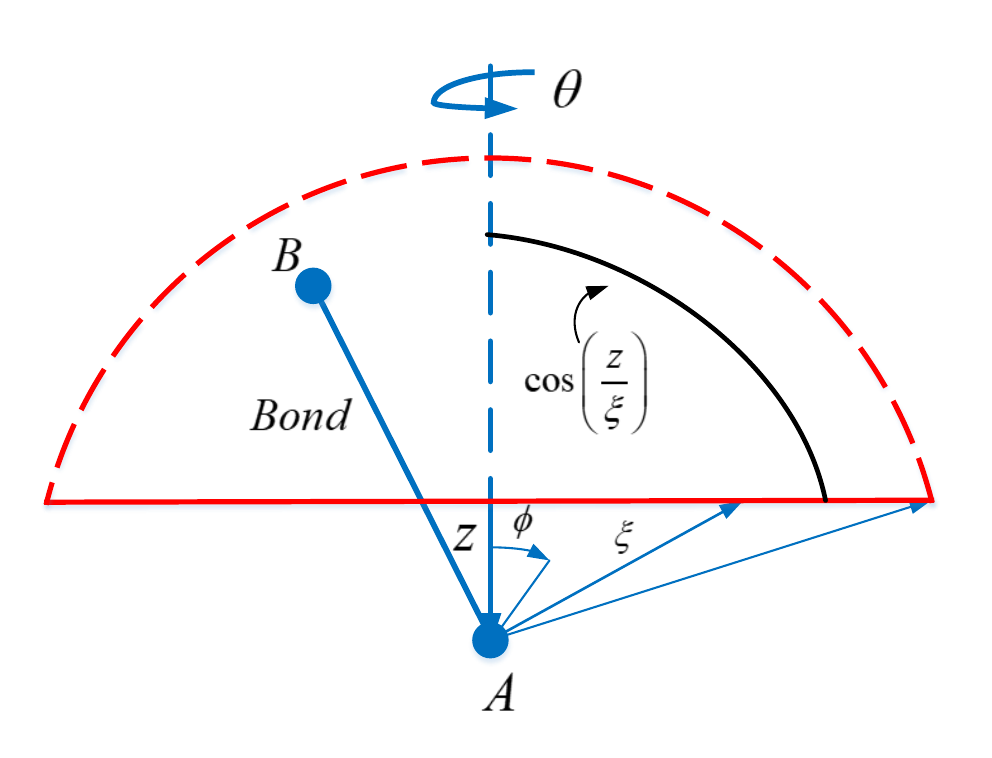}  
		\caption{Evaluation of fracture energy $G_0$. Here red line is the fracture surface.  $\mathbf{A}$ and  $\mathbf{B}$ are material points and  $0\leq z \leq \delta$.} 
		\label{Fig:s0}
	\end{figure}

 $s_c$ is a constant related to the work required to break a single bond. Here we use $w_0(\eta)$ denote the work, and can obtain
 \begin{equation}\label{w01}
 w_0(\eta)=\int_0^{s_c} g(s)d\eta=\int_0^{s_c} g(s)\xi ds=\dfrac{3E s_c^2G(\bm{\xi})\xi^4}{8\sigma^{4}}.
 \end{equation}
 Here, $G_0$ is introduced to represent the work required to break all the bonds per unit fracture area. For the 2D model, we have
 \begin{equation}\label{G01}
 G_0= 2h\int_0^{\infty} \int_z^{\infty} \int_0^{\cos ^{-1} z / \xi}\dfrac{3Es_c^2G(\bm{\xi})\xi^4}{8\sigma^{4}}\xi \mathrm{d} \phi \mathrm{d} \xi  \mathrm{d} z=\dfrac{45Ehs_c^2\sigma}{8\sqrt{2\pi}}.
 \end{equation}
 An explanation of this computation is shown in Fig. \ref{Fig:s0}. Assessed that our kernel function is built on an unbounded region, therefore the radius of action are $\infty$.
Then we can obtain
 \begin{equation}
G_0=\dfrac{45Ehs_c^2\sigma}{8\sqrt{2\pi}},
\end{equation}
thus
\begin{equation}\label{sc}
s_c=\sqrt{\dfrac{8\sqrt{2\pi}G_0}{45Eh\sigma}}.
\end{equation} 
\section{Linear two-dimension Gaussian-kernel bond-based peridynamics}
\label{section4}
In this section, a linear BBPD-GK is constructed.  Assume the displacement is small enough, which means $|\bm{\eta}|<1$, we can linearize the function  $\mathbf{g}(\bm{\xi},\bm{\eta})$ as follows:
\begin{equation}
    \mathbf{g}(\bm{\xi},\bm{\eta})=\mathbf{C}(\boldsymbol{\xi})\bm{\eta}.
\end{equation}
where $\mathbf{C}(\boldsymbol{\xi})$ is 
\begin{equation}\label{kerlin}
    \mathbf{C}(\boldsymbol{\xi})=\frac{\partial \mathbf{g}}{\partial \boldsymbol{\eta}}(\mathbf{0}, \boldsymbol{\xi}).
\end{equation}
Substitute \eqref{pde3} into \eqref{kerlin}, we can obtain
\begin{equation}
\mathbf{C}(\boldsymbol{\xi})=\dfrac{3E}{8\pi\sigma^6}exp\left({-\dfrac{|\bm{\xi}|}{2\sigma^2}}\right)\left(\bm{\xi}\otimes\bm{\xi}\right).
\end{equation}
Then, the linear BBPD-GK operation can be defined as
\begin{equation}\label{lin}
\mathcal{L}_{\sigma}^l\mathbf{u}(\mathbf{x},t)=\int_{\mathcal{R}}\mathbf{C}(\boldsymbol{\xi})\bm{\eta} dV_{\mathbf{\mathbf{x}^{\prime}}}+\mathbf{b}(\mathbf{x},t).
\end{equation}
and then the motion equation can be defined by
	\begin{equation}\label{pdelin}
	\rho \ddot{\mathbf{u}}(\mathbf{x},t)=\mathcal{L}_{\sigma}^l\mathbf{u}(\mathbf{x},t).
	\end{equation}

Suppose the solution $\mathbf{u}$ is sufficiently smooth. Consider a 2D model and  let $\mathbf{x}=(x,y)$, $\mathbf{u}(\mathbf{x},t)=(u(\mathbf{x},t),v(\mathbf{x},t))$, where $x$ and $y$ are components in the x- and y- directions, and $u(\mathbf{x},t)$ and $v(\mathbf{x},t)$ are displacement components. Using Taylor expansion, we have:
\begin{equation}\label{taylor}
\left[\begin{matrix}
u(\mathbf{\mathbf{x}^{\prime}},t)-u(\mathbf{\mathbf{x}},t)\vspace{0.5em}\\
v(\mathbf{\mathbf{x}^{\prime}},t)-v(\mathbf{\mathbf{x}},t)
\end{matrix}\right]=\left[\begin{matrix}
\nabla u^T(\mathbf{x},t)(\mathbf{x}^{\prime}-\mathbf{x})+\dfrac{1}{2}(\mathbf{x}^{\prime}-\mathbf{x})^T \mathbf{H}_u(\mathbf{x}^{\prime}-\mathbf{x})+\dots\vspace{0.5em}\\\nabla v^T(\mathbf{x},t)(\mathbf{x}^{\prime}-\mathbf{x})+\dfrac{1}{2}(\mathbf{x}^{\prime}-\mathbf{x})^T \mathbf{H}_v(\mathbf{x}^{\prime}-\mathbf{x})+\dots\\
\end{matrix}\right]
\end{equation}
with $\mathbf{H}_u$ and $\mathbf{H}_v$ are the Hessians of $u$ and $v$, respectively. When Substituting \eqref{taylor}  into \eqref{pdelin}, according to the definition of three-order moments for the multivariate Gaussian distribution \cite{moment}, it can be observed that the integrals of the first  terms  are $0$, which means
\begin{align}
&\int_\mathcal{R} \dfrac{1}{2\pi\sigma^2}exp\left({-\dfrac{\xi^2}{2\sigma^2}}\right)\left(x^{\prime}-x\right)^3dV_{\mathbf{\mathbf{x}^{\prime}}}=0,\notag\\
&\int_\mathcal{R} \dfrac{1}{2\pi\sigma^2}exp\left({-\dfrac{\xi^2}{2\sigma^2}}\right)\left(x^{\prime}-x\right)^2\left(y^{\prime}-y\right)dV_{\mathbf{\mathbf{x}^{\prime}}}=0,\notag\\
&\int_\mathcal{R} \dfrac{1}{2\pi\sigma^2}exp\left({-\dfrac{\xi^2}{2\sigma^2}}\right)\left(x^{\prime}-x\right)\left(y^{\prime}-y\right)^2dV_{\mathbf{\mathbf{x}^{\prime}}}=0,\\
&\int_\mathcal{R} \dfrac{1}{2\pi\sigma^2}exp\left({-\dfrac{\xi^2}{2\sigma^2}}\right)\left(y^{\prime}-y\right)^3dV_{\mathbf{\mathbf{x}^{\prime}}}=0,\notag
\end{align}
then, according to the definition of fourth-order moments for the multivariate Gaussian distribution \cite{moment},  we have the following equation
\begin{align}\label{moment}
& \int_\mathcal{R} \dfrac{1}{2\pi\sigma^2}exp\left({-\dfrac{\xi^2}{2\sigma^2}}\right)\left(x^{\prime}-x\right)^4dV_{\mathbf{\mathbf{x}^{\prime}}}=3\sigma^{4},\notag\\
&\int_\mathcal{R} \dfrac{1}{2\pi\sigma^2}exp\left({-\dfrac{\xi^2}{2\sigma^2}}\right)\left(x^{\prime}-x\right)^3\left(y^{\prime}-y\right)dV_{\mathbf{\mathbf{x}^{\prime}}}=0,\notag\\
&\int_\mathcal{R} \dfrac{1}{2\pi\sigma^2}exp\left({-\dfrac{\xi^2}{2\sigma^2}}\right) \left(x^{\prime}-x\right)^2\left(y^{\prime}-y\right)^2dV_{\mathbf{\mathbf{x}^{\prime}}}=\sigma^{4},\\
& \int_\mathcal{R} \dfrac{1}{2\pi\sigma^2}exp\left({-\dfrac{\xi^2}{2\sigma^2}}\right) \left(x^{\prime}-x\right)\left(y^{\prime}-y\right)^3dV_{\mathbf{\mathbf{x}^{\prime}}}=0,\notag\\
&\int_\mathcal{R} \dfrac{1}{2\pi\sigma^2}exp\left({-\dfrac{\xi^2}{2\sigma^2}}\right) \left(y^{\prime}-y\right)^4dV_{\mathbf{\mathbf{x}^{\prime}}}=3\sigma^{4},\notag
\end{align}
Substitute \eqref{moment} into \eqref{taylor}, we can obtain 
\begin{equation}\label{conver}
	\mathcal{L}_{\sigma}^l\mathbf{u}(\mathbf{x},t)=\left[\begin{matrix}
	\dfrac{9E}{8}\dfrac{\partial^2 u(\mathbf{x},t)}{\partial x^2}+\dfrac{3E}{4}\dfrac{\partial^2 v(\mathbf{x},t)}{\partial x\partial y}+\dfrac{3E}{8}\dfrac{\partial^2 u(\mathbf{x},t)}{\partial y^2}\vspace{0.5em}\\\dfrac{9E}{8}\dfrac{\partial^2 v(\mathbf{x},t)}{\partial x^2}+\dfrac{3E}{4}\dfrac{\partial^2 u(\mathbf{x},t)}{\partial x\partial y}+\dfrac{3E}{8}\dfrac{\partial^2 v(\mathbf{x},t)}{\partial y^2}
	\end{matrix}\right],
\end{equation}
which can be expressed as
\begin{equation}
	\mathcal{L}_{\sigma}^l\mathbf{u}(\mathbf{x},t)\rightarrow\nabla\cdot(\mathbf{D}:\bm{\varepsilon}).
\end{equation}
Here $\nu$ is Poisson's ratio and is chosen as $1/3$ for two-dimension model, and the matrix of isotropic elastic moduli takes the usual form
\begin{equation}
\mathbf{D}= \frac{E}{1-v^2}\left[\begin{array}{ccc}
 1 & v & 0 \\
 v & 1 & 0 \\
 0 & 0 & \frac{1}{2}(1-v)
 \end{array}\right]
\end{equation}
and $\bm{\varepsilon}$ is the strain tensor which can be defined by
\begin{equation}
    \bm{\varepsilon}=\dfrac{\nabla\mathbf{u}(\mathbf{x},t)+\nabla\mathbf{u}^T(\mathbf{x},t)}{2}.
\end{equation}
It can be observed that the right-hand side of \eqref{conver} is the motion equation form of classical elasticity, which means that  the proposed BBPD-GK can be compatiable to the form of classical elasticity. 
\section{Three-dimension Gaussian-kernel bond-based peridynamics}
\label{section5}
In this section, we discuss the three-dimensional BBPD-GK model. Similar to a two-dimensional form, we introduce a three-dimensional  motion equation as follows
\begin{equation}
\rho \ddot{\mathbf{u}}(\mathbf{x},t)=\int_{\mathcal{R}}\dfrac{\boldsymbol{\xi}+\boldsymbol{\eta}}{|\boldsymbol{\xi}+\boldsymbol{\eta}|}\beta s \dfrac{1}{\sqrt{(2\pi\sigma^2)^3}}exp\left({-\dfrac{|\bm{\xi}|}{2\sigma^2}}\right)|\bm{\xi}|^3 dV_{\mathbf{\mathbf{x}^{\prime}}}+\mathbf{b}(\mathbf{x},t),
\end{equation}
and according to \cite{PD}, for 3D model, we have 
  \begin{equation}
  	W=\frac{1}{2}\int_\mathcal{R} w dV_{\mathbf{\mathbf{x}^{\prime}}}=\dfrac{1}{2} \int_0^\infty \dfrac{\beta s^2 \xi^4 G(\bm{\xi})}{2}4 \pi \xi^2 d\xi=\dfrac{15}{4}\beta s^2 \sigma^4,
  \end{equation}
  thus
   \begin{equation}
    \beta= \dfrac{4E}{5\sigma^{4}}.
 \end{equation}
 Similar to 2D model, we can establish the expression of the critical stretch  
  \begin{equation}\label{gnolin2}
 	\quad \mu= \begin{cases} 1, &\text{ $s<s_c$}, \\ 0, & \text { $s \ge s_c$},\end{cases}
 \end{equation}
 where $s_c$ can be computed by
 \begin{equation}
 	G_0= \int_0^{\delta} \int_0^{2\pi}\int_z^{\delta} \int_0^{\cos ^{-1} z / \xi}\dfrac{4E s_c^2G(\bm{\xi})\xi^4}{10\sigma^{4}}\xi^3sin\phi \mathrm{d} \phi \mathrm{d} \xi \mathrm{d} \theta \mathrm{d} z=\dfrac{48Es_c^2\sigma}{5\sqrt{2\pi}},
 \end{equation}
 then we can obtain
 \begin{equation}
s_c=
\sqrt{\dfrac{5\sqrt{2\pi}G_0}{48E\sigma}},
\end{equation}
With sufficiently small displacement, the model can be approximated using linear BB-PD peridynamics, which is
    	\begin{equation}
	\rho \ddot{\mathbf{u}}(\mathbf{x},t)=\int_{\mathcal{R}}\dfrac{4E}{5\sqrt{8\pi^3}\sigma^{7}}exp\left({-\dfrac{|\bm{\xi}|}{2\sigma^2}}\right)\left(\bm{\xi}\otimes\bm{\xi}\right)\bm{\eta} dV_{\mathbf{\mathbf{x}^{\prime}}}+\mathbf{b}(\mathbf{x},t).
	\end{equation} 
 
\section{Numerical discretization}  
\label{section6}
\subsection{Truncation of the influence region for the kernel function}
The proposed BBPD-GK uses unbounded functions instead of bounded functions in BB-PD, which poses computational difficulties. However, due to the fast decay of the selected kernel function $G(\bm{\xi})$, we can use a truncated bounded region instead of an unbounded region. By providing an appropriate cutoff distance to control the truncation region, the computational burden of the model can be reduced without affecting its computational accuracy, thereby increasing the practicality of the model.

According to \cite{moment}, $\bm{\xi}^2$ obey the chi-square distribution when $\bm{\xi}$ satisfy the d-dimensional Gaussian distribution $\chi^2(d)$. Thus, we can take the influence region of the kernel function at $\mathbf{x}$ which satisfies $\bm{\xi}^2\leq\sigma^2\chi_a^2(d)$ for all $\mathbf{x}^{\prime}$, where  $\chi_a^2(d)$  is the $(1-a)$  quantile of the chi-square distribution and $a$ is a predefined constant. Then we have
\begin{equation}
\int_{\bm{\xi}^2 \leq \delta^2 \chi_a^2(d)} G (\bm{\xi}) d \mathbf{x}^{\prime}=1-a
\end{equation}
We will take
\begin{equation}
B_{\sigma, \alpha}(\boldsymbol{x})=\left\{\boldsymbol{x}^{\prime}\mid \bm{\xi}^2\leq\sigma^2\chi_a^2(d)\right\},\quad 0\le\alpha\leq 1,
\end{equation}
as the truncated influence region for $\bm{x}$ with $a$ selected very close to $0$.  By introducing the truncated influence region, the force density force $\mathbf{f}$ can be rewritten as follows
   \begin{equation}\label{f2}
g(\bm{\eta},\bm{\xi})=\left\{\begin{array}{lll}
\beta sG(\bm{\xi}),\quad&\mathbf{x}^{\prime}\in B_{\sigma, a}(\boldsymbol{x})  \text { and } s<s_{0} \\
0, \quad&\mathbf{x}^{\prime}\notin B_{\sigma, a}(\boldsymbol{x})\text { or } s \geq s_{0}
\end{array}\right.
\end{equation}

\subsection{Meshfree method}

In order to numerically simulate the proposed BBPD-GK, we discretize it by following the meshfree approach proposed in \cite{SE}. In meshfree method, the region is discretized into nodes $\mathbf{x}_i$, each with a known volume $\Delta x_i$ in the reference configuration, $i=1,2\dots N$, where  $N$ is the total number of the material points in the truncation of the influence region. The discretized form of the nonlinear  motion equation \eqref{pde3} replaces the integral by a finite sum:
\begin{equation}
	\rho \ddot{\mathbf{u}}_i^n=\sum\limits_{j=1}^N \mathbf{g}\left(\mathbf{u}(\mathbf{x}_j,t)-\mathbf{u}(\mathbf{x}_i,t), \mathbf{x}_j-\mathbf{x}_i\right) V_j+\mathbf{b}_i^n,
\end{equation}
where $\mathbf{g}$ is supplied by \eqref{pde3}. Similarly, a finite sum for the linear model obtained by the meshfree method can be expressed as 
\begin{equation}
	\rho \ddot{\mathbf{u}}_i^n=\sum\limits_{j=1}^N \mathbf{g}\left(\mathbf{u}(\mathbf{x}_j,t)-\mathbf{u}(\mathbf{x}_i,t), \mathbf{x}_j-\mathbf{x}_i\right) V_j+\mathbf{b}_i^n,
\end{equation}

\begin{figure}[ht]
	\centering    	
	\subfigure[]
	{
		\begin{minipage}{7cm}
			\centering          
			\includegraphics[scale=0.2]{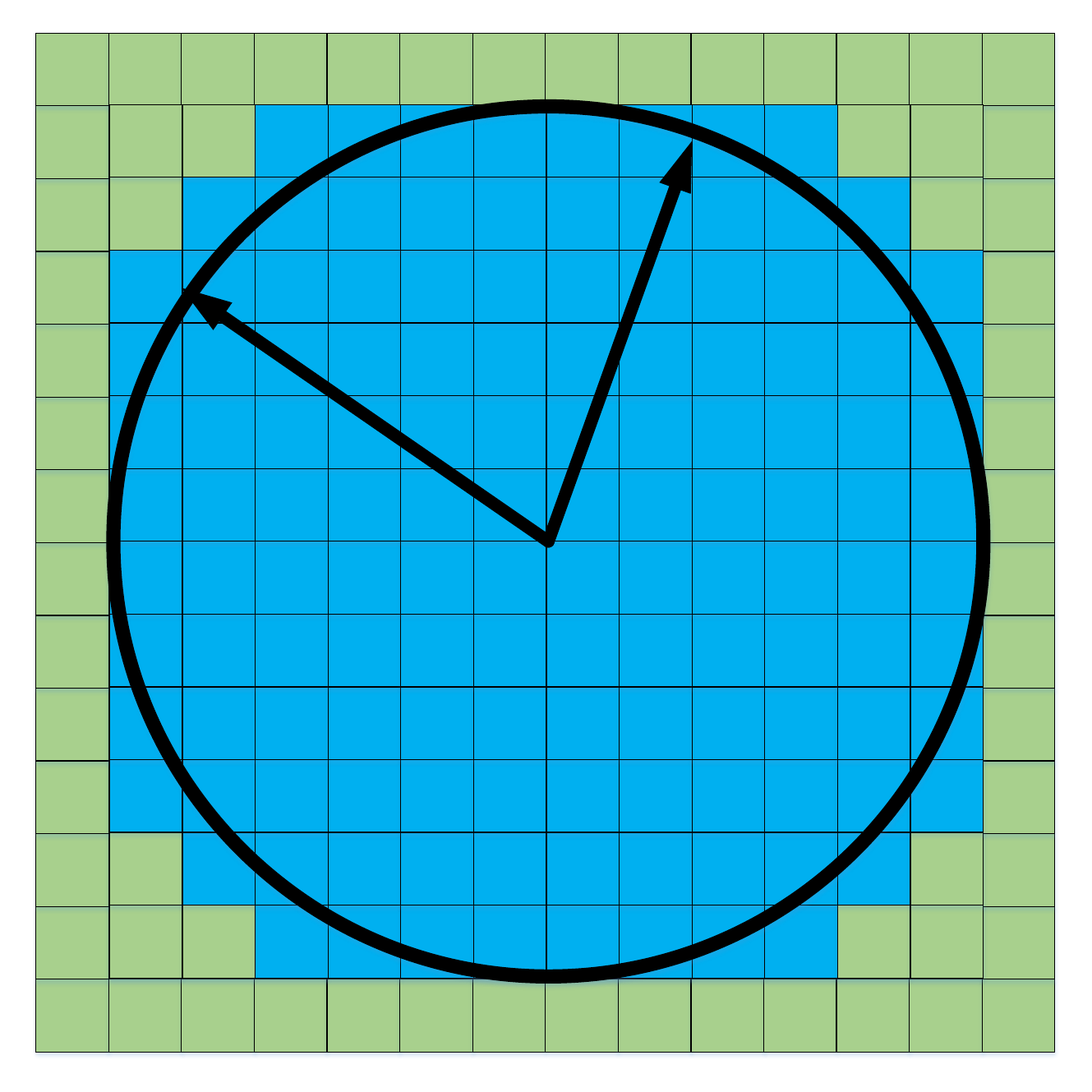}  
		\end{minipage}
	}	
	\subfigure[]
	{
		\begin{minipage}{7cm}
			\centering      
			\includegraphics[scale=0.15]{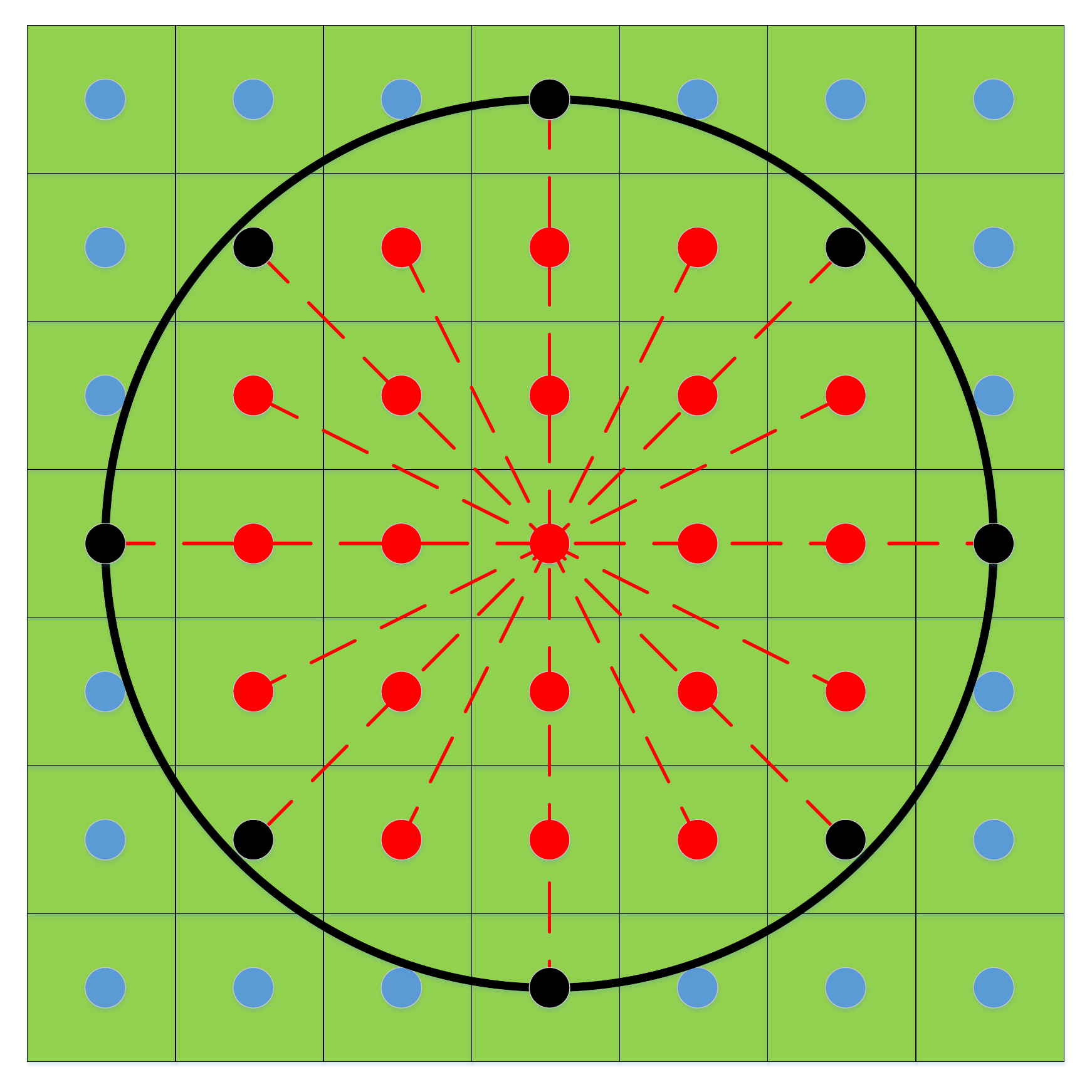}   
		\end{minipage}
	}	
	\caption{ Two models discretized by the meshfree method: (a) Truncation of the influence region with the $\chi_a^2(2)=36$ and $\sigma=\Delta x$; (b) The volume correction in PD with horizon size $\delta=3h$.
} 
	\label{Fig:3}
\end{figure}

One important consideration is that while we initially truncated the unbounded region using a circular area, challenges arose when adopting the meshless method for discretization. Specifically, difficulties emerged in calculating the volume of the intersecting region between a PD point and its neighboring cells. Consequently, we revised the circular region. As depicted in the Fig. \ref{Fig:3} (a), if the material points within the truncated area meet the condition where only a portion of the cell intersects the horizon, we calculate the volume of the entire grid. It's essential to note that the truncation region is based on estimating unbounded regions using a normal distribution, rather than actual calculation. Therefore, expanding the truncation region does not compromise calculation accuracy.

\begin{remark}
	In contrast to the proposed Gaussian-type kernel-based peridynamics, many existing
	peridynamics restrict nonlocal interactions to bounded neighborhoods, often chosen as Euclidean disks or balls. The approximate disks or balls, typically composed of polygons, impose a challenge when intersecting for meshfree discretization method. To address this issue, as shown in Fig. \ref{Fig:3} (b), a volume correction parameter $\lambda_{i,j}$ is introduced, namely
	\begin{equation}\label{mat:f1}
	\rho\ddot{\mathbf{u}}_{i}=\sum_{\x_j\in\mathcal{R}} \lambda_{i,j}\mathbf{f}\left(\mathbf{u}(\mathbf{x}_j,t)-\mathbf{u}(\mathbf{x}_i,t), \mathbf{x}_j-\mathbf{x}_i\right) V_{j},
	\end{equation}
   As shown in the Fig. \ref{Fig:3} (b), the value of $\lambda_{i,j}$ will also vary depending on the intersection between the volume of the material point and the horizon.  A lot of volume correction methods have been developed to give a proper choice of $\lambda_{i,j}$, including FA method \cite{MF}, LAMMA method \cite{FAC}, and QWJ method \cite{vol3}. More detailed description of these method are presented in Section 8.2.
\end{remark}

   Regarding the asymptotic compatibility, we will provide numerical evidence in the next section to illustrate that meshfree discretization achieves
   $\sigma$-convergence. Notably, we will demonstrate that the numerical solutions of the nonlocal model under a fixed ratio between $\sigma$ and $\Delta x$ exhibit convergence towards the corresponding local PDE solution.
\section{Validations}
\label{section7}
The applicability of the proposed BBPD-GK model, including the convergence, accuracy, and rationality of model parameter selection, was demonstrated through a verification example of a two-dimensional plate with displacement constraints.

\begin{figure}[ht]
		\centering            
		\includegraphics[scale=0.45]{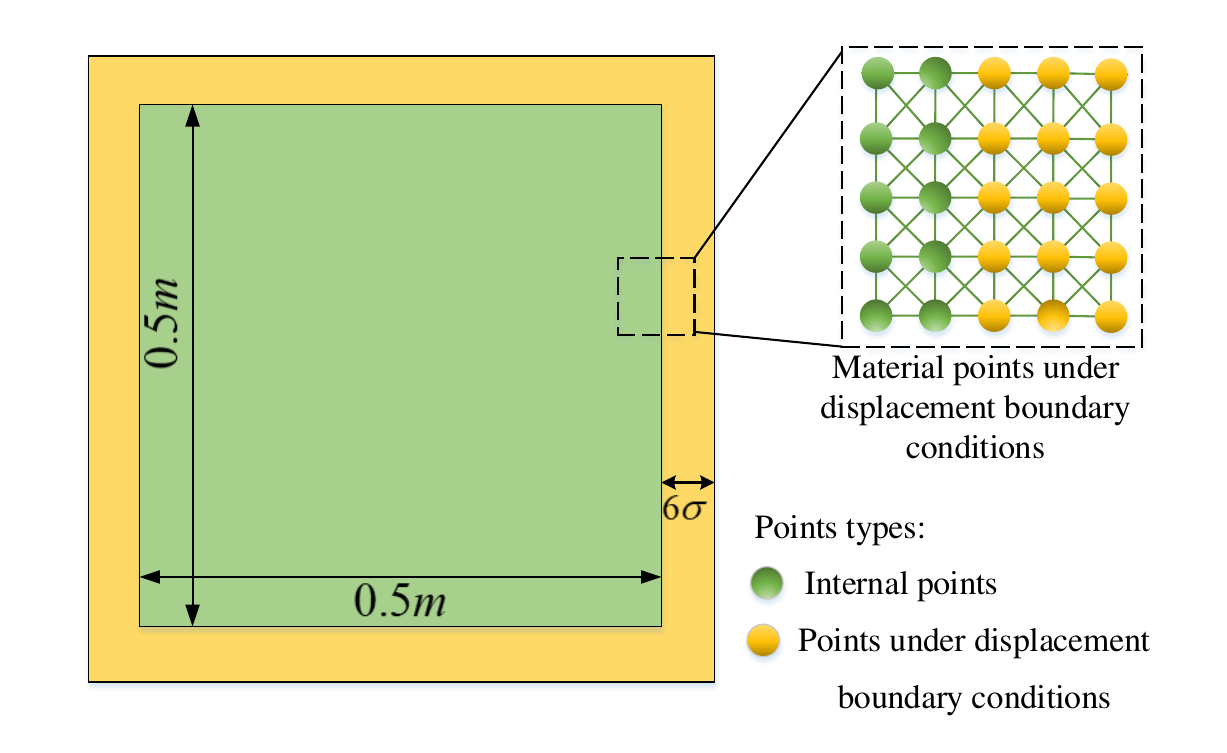}  
		\caption{Geometry of a plate with displacement boundary condition.} 
		\label{Fig:val}
	\end{figure}

Here, we test the peridynamic solution with variable models for the quasi-static stretching of an elastic plate.
As shown in Fig. \ref{Fig:val},  this plate has a length $L=0.5\mathrm{~m}$, width $W=0.5\mathrm{~m}$, and thickness $h=0.0025\mathrm{~m}$. Microelastic models corresponding to a linear elastic material with $E=1.92\times 10^5$ Mpa (elastic modulus), $v=1/3$ (Poisson's ratio) and $\rho=8000 \mathrm{~kg}/\mathrm{m}^3$ (density) are used in all models below. Let $\mathbf{u}_h(\mathbf{x})=(u_h(\mathbf{x}),v_h(\mathbf{x}))$, $\mathbf{x}=(x,y)$ be the reference solution for the problem, which can be expressed as $u_h=sinxy$, $v_h=cosxy$. Without losing generality, the displacement constraints $\mathbf{u}(\mathbf{x})=\mathbf{u}_h(\mathbf{x})$ is applied on the boundary zone surrounding the plate with a width 
$6\sigma$.

\subsection{Tests for asymptotically compatible convergence}
 To verify the convergence of the proposed BBPD-GK, In the following, we uniformly partition the domain to $N\times N$ rectangle cells, where $N=51,101,201$. The grid spacing is clearly the same for both $x-$ and $y-$ directions. 
 \begin{table}[!ht]\begin{center}
		\caption{The  $L^2$ solution errors $e$ and convergence rates CR
computed by linear BBPD-GK with different $\sigma$.}
		\vspace{0.15in}
		\setlength{\tabcolsep}{4mm}{
			\begin{tabular}{|c|c|c|c|c|c|c|} \hline
				&\multicolumn{2}{|c|}{$\sigma=\Delta x/2$}&\multicolumn{2}{c|}{$\sigma=\Delta x$}&\multicolumn{2}{c|}{$\sigma=2\Delta x$}\\ \hline
				 N&$e$ & CR&$e$ & CR& $e$& CR\\ \hline				
				51&$4.87$E$-5$&-&$8.73$E$-8$&-&$3.89$E$-7$&- \\
				101&$4.63$E$-5$&0.07&$2.05$E$-8$&2.09&$8.52$E$-8$&2.18\\
				201&$4.51$E$-5$&0.04&$5.43$E$-9$&1.92&$4.88$E$-9$&2.02\\\hline				
		\end{tabular}}\label{tab:1}
\end{center}\end{table}

Table \ref{tab:1} reports the discrete $L^2$ solution errors and corresponding convergence rates produced
by the linear model noted in \eqref{pdelin} with  $\sigma=\Delta x/2, \Delta x, 2\Delta x$. It can be observed that when $\sigma=\Delta x, 2\Delta x$,  the BBPD-GK reach their respective optimal temporal second-order convergence along the grid refinement. When $\sigma=0.5\Delta x$, the results fail to be convergent.  This implies that similar to the nonlocal neighborhood radius $\delta$ in BB-PD, $\sigma$ is a coefficient related to $\Delta x$, and accuracy improves as the ratio $\sigma/\Delta x$ increases. IIt should be noted that when $\sigma$ is chosen as $\Delta x$, BB-GKPD achieves the expected convergence order. Therefore, considering computational cost, variance is defined as $\sigma=\Delta x$ to avoid the computational burden of selecting larger values in the following simulation, such as $\sigma=2\Delta x$.
 \begin{table}[!ht]\begin{center}
		\caption{The  $L^2$ solution errors $e$ and convergence rates CR
computed by linear BBPD-GK with different $\chi_a^2(2)$.}
		\vspace{0.15in}
		\setlength{\tabcolsep}{2mm}{
			\begin{tabular}{|c|c|c|c|c|c|c|c|c|c|c|} \hline
				&\multicolumn{2}{|c|}{$\chi_a^2(2)=16$}&\multicolumn{2}{c|}{$\chi_a^2(2)=25$}&\multicolumn{2}{c|}{$\chi_a^2(2)=36$}&\multicolumn{2}{c|}{$\chi_a^2(2)=49$}\\ \hline
				 N&$e$ & CR& $e$ & CR& $e$ & CR& $e$ & CR\\ \hline				
				51&$6.39$E$-6$&-&$1.55$E$-7$&-&$8.73$E$-8$&-&$8.67$E$-8$&- \\
				101&$5.88$E$-6$&0.12&$8.36$E$-8$&0.89&$2.05$E$-8$&2.09&$1.99$E$-8$&2.12\\
				201&$5.64$E$-6$&0.06&$6.62$E$-8$&0.33&$5.43$E$-9$&1.92&$4.88$E$-9$&2.02\\\hline				
		\end{tabular}}\label{tab:2}
\end{center}\end{table}

Another noteworthy question is the selection of truncated regions. For comparison purposes,  Specifically, we let $\chi_a^2(2)=16,25,36,49,64$,  respectively, and keep the remaining parameters unchanged. The $L^2$ solution errors computed by the linear model are shown in Table \ref{tab:2}. We can observe that when $\chi_a^2(2)=16,25,36$, the error gradually decreases while the order steadily increases. The errors between $\chi_a^2(2)=36$ and $\chi_a^2(2)=49$ are almost similar and In these two examples, ideal convergence results were achieved. Therefore, in the following experiment, we chose $\chi_a^2(2)=36$ to ensure the accuracy of the model and avoid additional computational costs.
\subsection{Tests for volume correction}
In this subsection, we consider comparing the accuracy of BBPD-GK and BB-PD with volume correction methods. Here, three volume correction method is introduced to compare with the BBPD-GK, one is  the FA method, which set 
\begin{equation}\label{fac:e0}
\hspace{-8em}
	\lambda_{p,q}=
	\begin{cases}
	1,& \text{if $|\bm{\xi}|\leq \delta$}, \\
	0,&\text{otherwise},
	\end{cases}
	\end{equation}
LAMMPS method and QWJ method are two other common correction methods, which for LAMMPS method, $\lambda_{p,q}$ can be expressed as
	\begin{equation}\label{fac:e1}
	\lambda_{p,q}=
	\begin{cases}
	1,& \text{if $|\bm{\xi}|\leq \delta-\dfrac{h}{2}$}, \\
	\dfrac{\delta-|\bm{\xi}|}{h}+\dfrac{1}{2},&\text{if $\delta-\dfrac{h}{2}<|\bm{\xi}|\leq \delta$}, \\
	0,&\text{otherwise},
	\end{cases}
	\end{equation}
and for the QWJ method, we have
	\begin{equation}\label{fac:e2}
 \hspace{1em}
	\lambda_{p,q}=
	\begin{cases}
	1,& \text{if $|\bm{\xi}|\leq \delta-\dfrac{h}{2}$}, \\
	\dfrac{\delta-|\bm{\xi}|}{h}+\dfrac{1}{2},&\text{if $\delta-\dfrac{h}{2}<|\bm{\xi}|\leq \delta+\dfrac{h}{2}$}. \\
	0,&\text{otherwise},
	\end{cases}
	\end{equation}

 We need to maintain generality by considering four different models for this plate. One is the linear BBPD-GK noticed in \eqref{pdelin}, and the other is the BB-PD with three volume correction parameters. We take a uniform grid of $N\times N$ with $N=51,101,201$ for all models and choose $\sigma=\Delta x$, 
$\chi_a^2(2)=36$ for BBPD-GK, $\delta=6\Delta x$ for BB-PD.

\begin{figure}[ht]
	\centering    	
	\subfigure[]
	{
		\begin{minipage}{7cm}
			\centering          
			\includegraphics[scale=0.3]{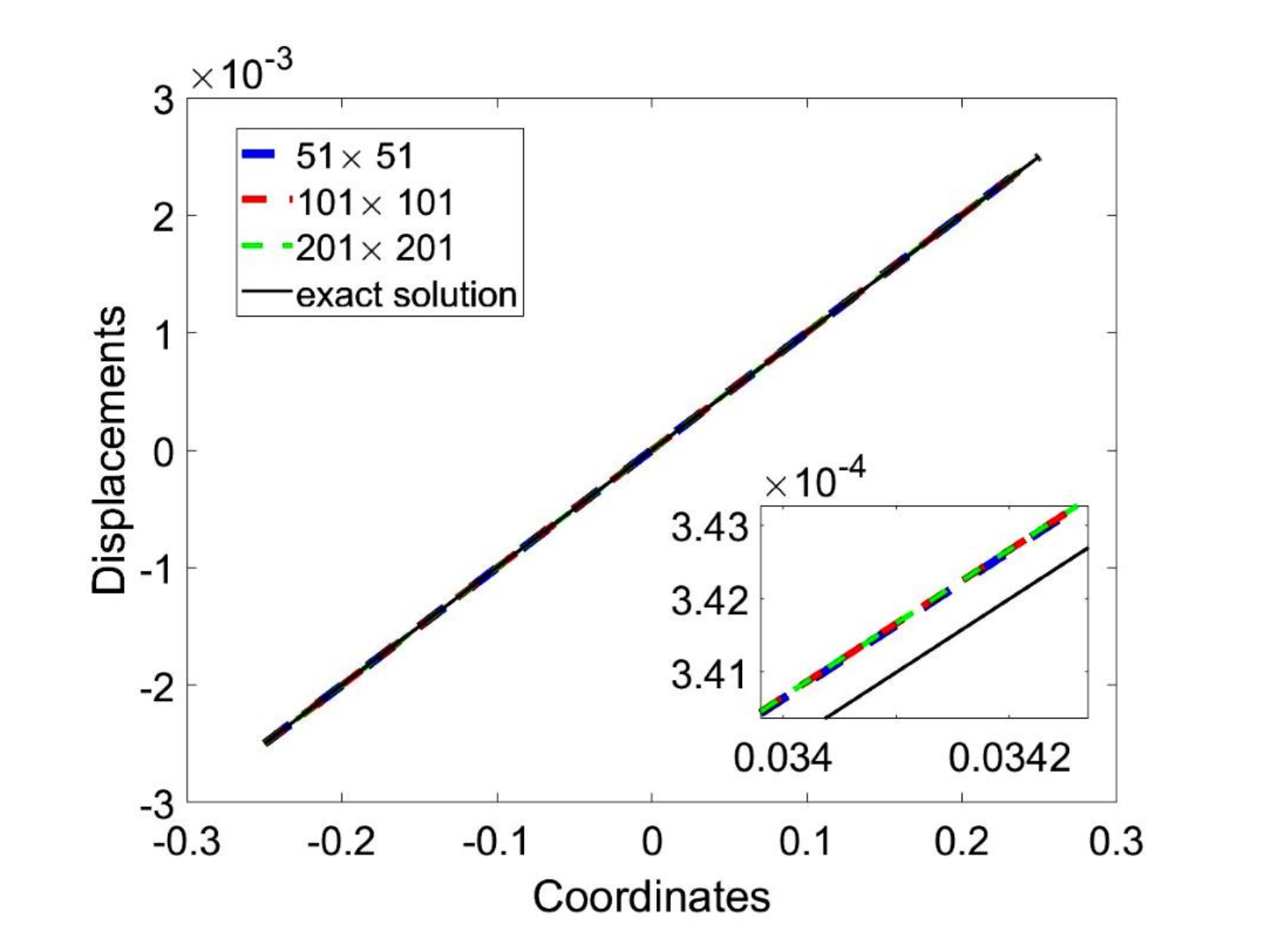}  
		\end{minipage}
	}	
	\subfigure[]
	{
		\begin{minipage}{7cm}
			\centering      
			\includegraphics[scale=0.3]{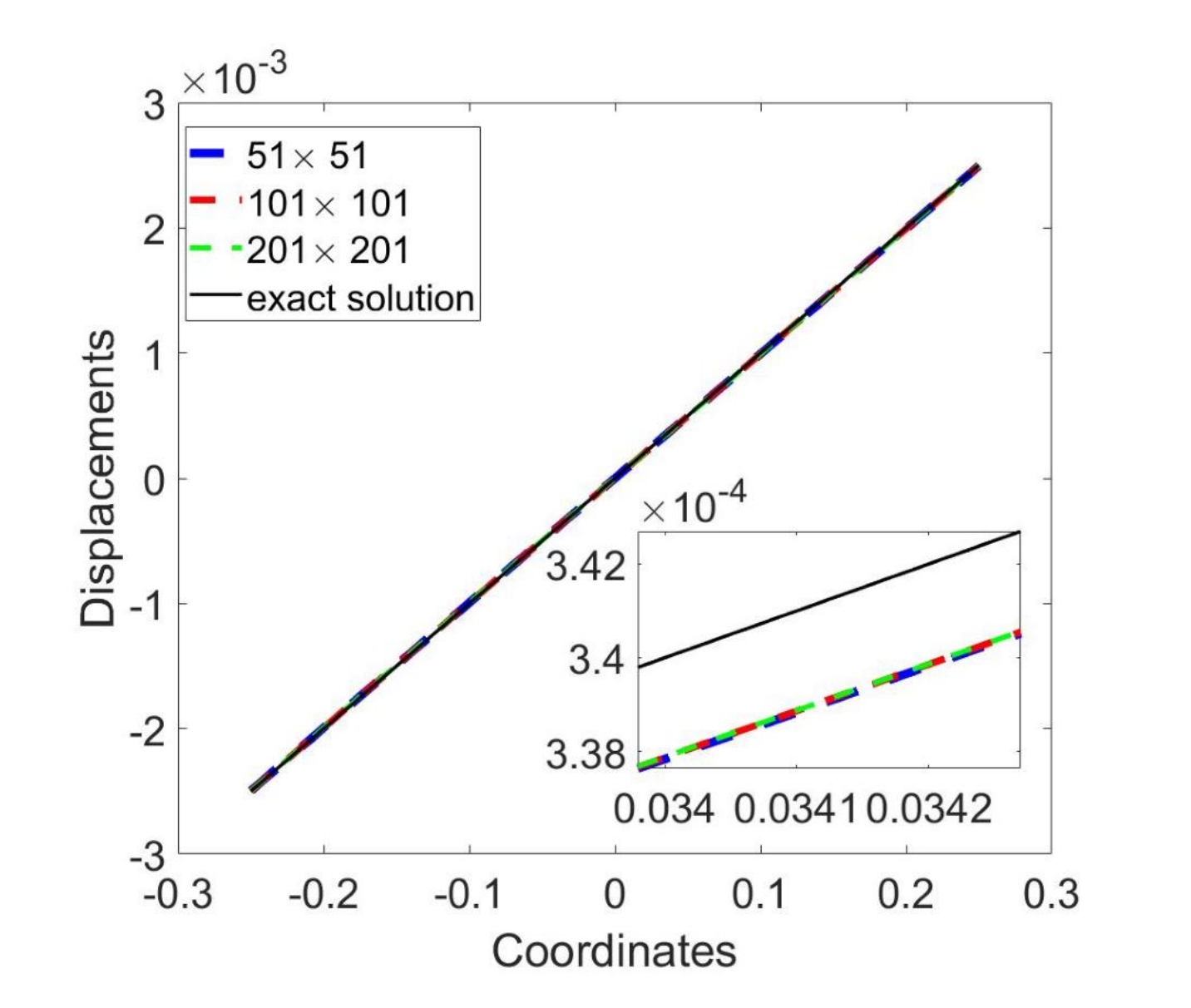}   
		\end{minipage}
	}	
 	\subfigure[]
	{
		\begin{minipage}{7cm}
			\centering          
			\includegraphics[scale=0.18]{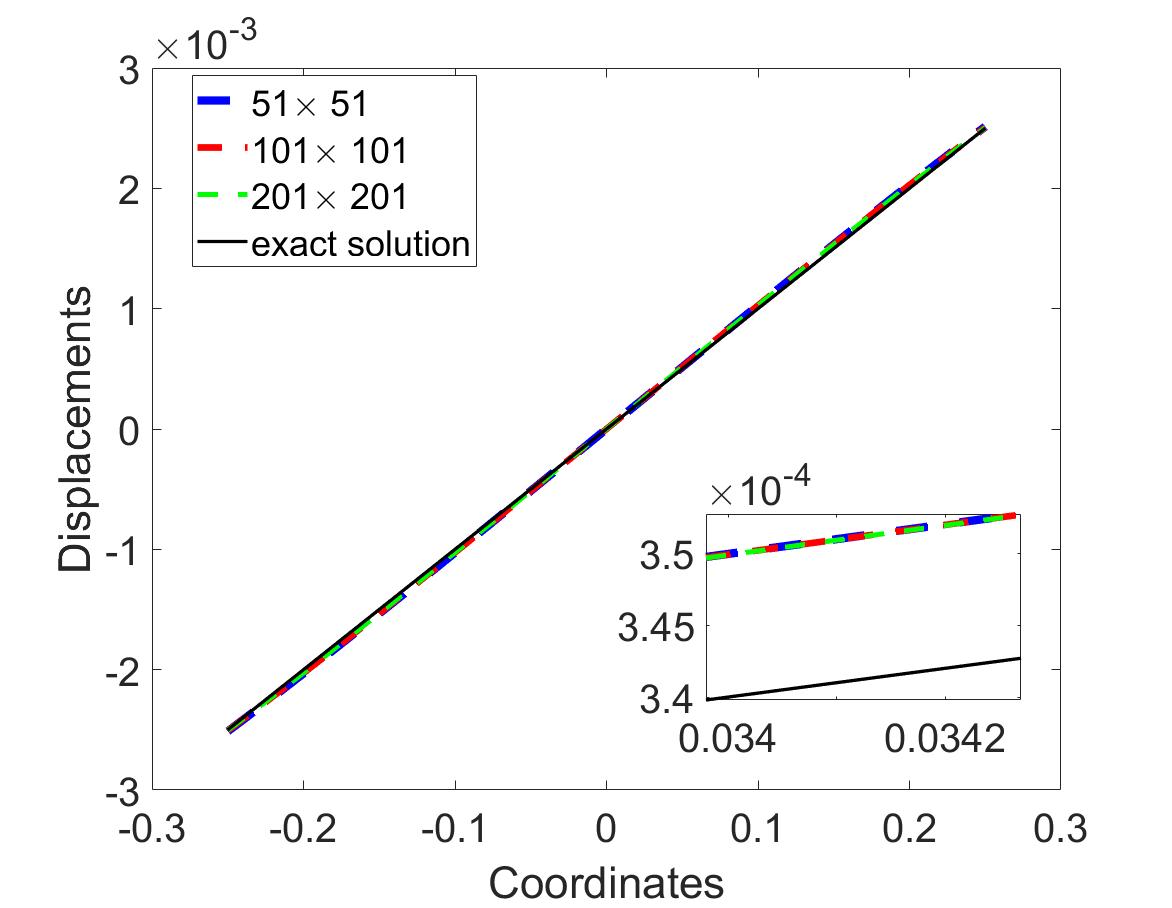}  
		\end{minipage}
	}	
	\subfigure[]
	{
		\begin{minipage}{7cm}
			\centering      
			\includegraphics[scale=0.18]{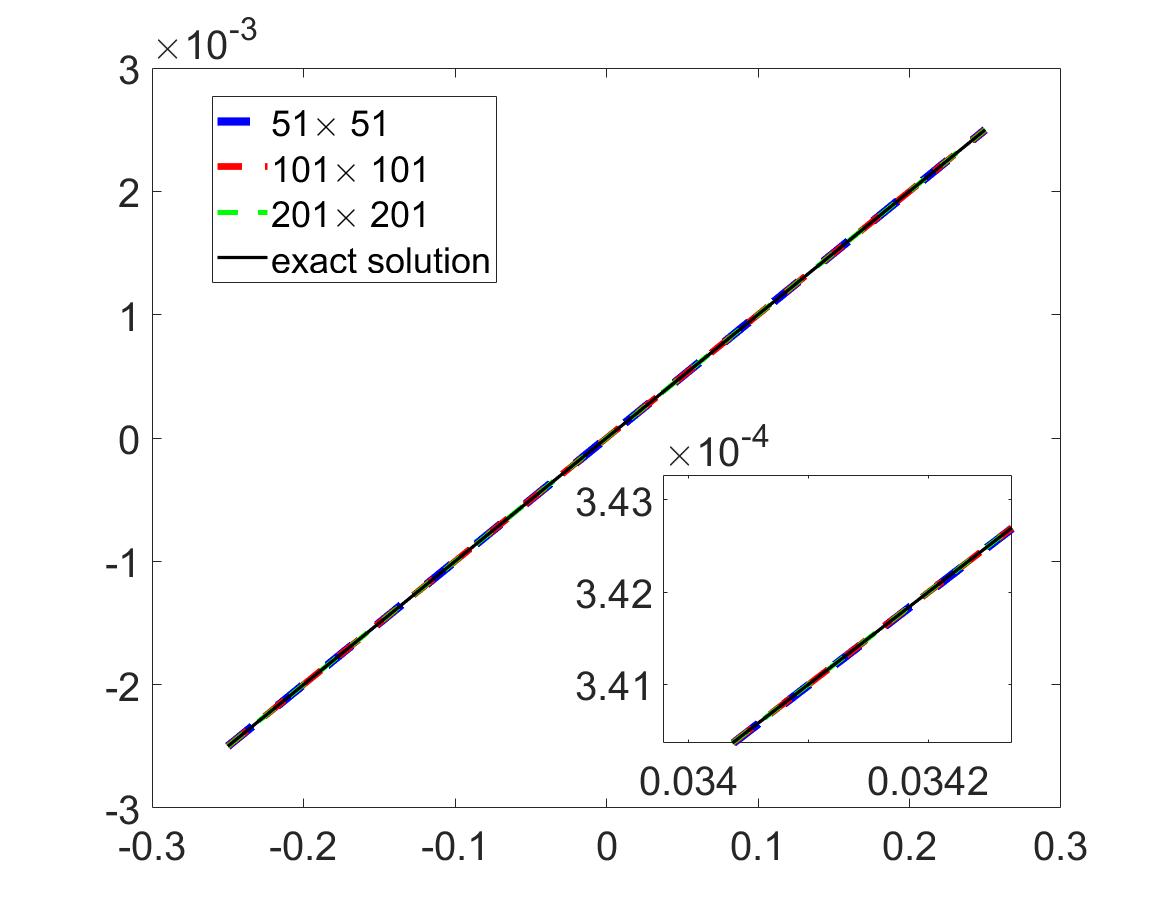}   
		\end{minipage}
	}	
	\caption{Displacement variation in y-direction when $x=0$: (a)By BB-PD with FA method; (b)By BB-PD with LAMMA method; (c) By BB-PD with QWJ method; (d) By BBPD-GK with no volume correction.
} 
	\label{Fig:8.5}
\end{figure}

 Fig. \ref{Fig:8.5} shows the displacement variation in the y-direction when $x=0$. It can be observed in Fig. that all load-displacement curves from our BBPD-GK and exact solution are well-aligned,  especially when the number of grids increases. The displacement obtained by BB-PD with three volume correct parameters fails to be compatiable to the exact solution, even the uniform grids are chosen as $1/401$, as shown in Fig. \ref{Fig:8.5} (a), (b) and (c).  This means that our proposed algorithm can achieve AC convergence through direct discretization. As the number of grids increases, the numerical solution will gradually approach the exact solution, a feat that traditional BB-PD models cannot achieve. 

\begin{figure}[ht]
		\centering            
		\includegraphics[scale=0.35]{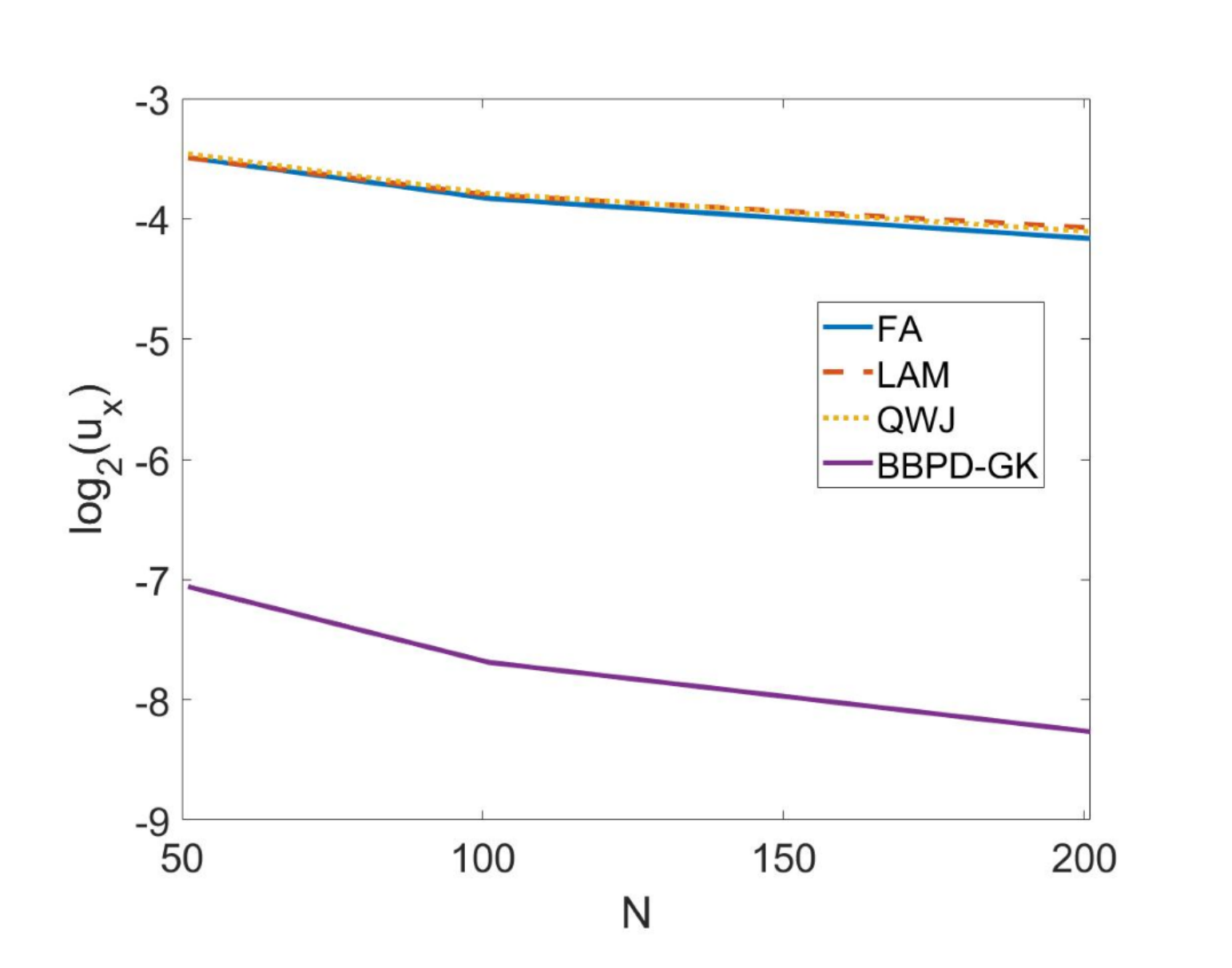}  
		\caption{The comparison of numerical error between BB-PD model with three different volume correction methods and BBPD-GK without volume correction} 
		\label{Fig:L2}
	\end{figure}

To quantitatively measure the accuracy of BBPD-GK, the $L^2$  error is calculated at different uniform girds by calculating the difference between displacement obtained by four models and the exact solution. As depicted in Fig. \ref{Fig:L2}, compared to the FA method, the QWJ method, and the LAMMPA method,  the BBPD-GK model has achieved higher accuracy. For example, the minimum error of the LAMMPA methods occurs when uniform girds are chosen as $1/201$, which is $8.43\times 10^{-5}$, and the error of BBPD-GK is $8.63\times 10^{-8}$ when grids are chosen as $1/401$. This means that our proposed model is capable of dealing with the problem of accurate evaluation of intersecting for meshfree discretization method and producing similar results regardless of the number of particles.
 \section{Numerical examples}
 \label{section8}
This section verifies two-dimension and three-dimension models with various boundary conditions and cracks. We implement these methods in Matlab and run all experiments on a workstation with Intel Xeon Gold 6240 (2.6GHz) logical processors and 2048G installed memory.
\subsection{Plate with a pre-existing crack}
\begin{figure}[ht]
		\centering            
		\includegraphics[scale=0.45]{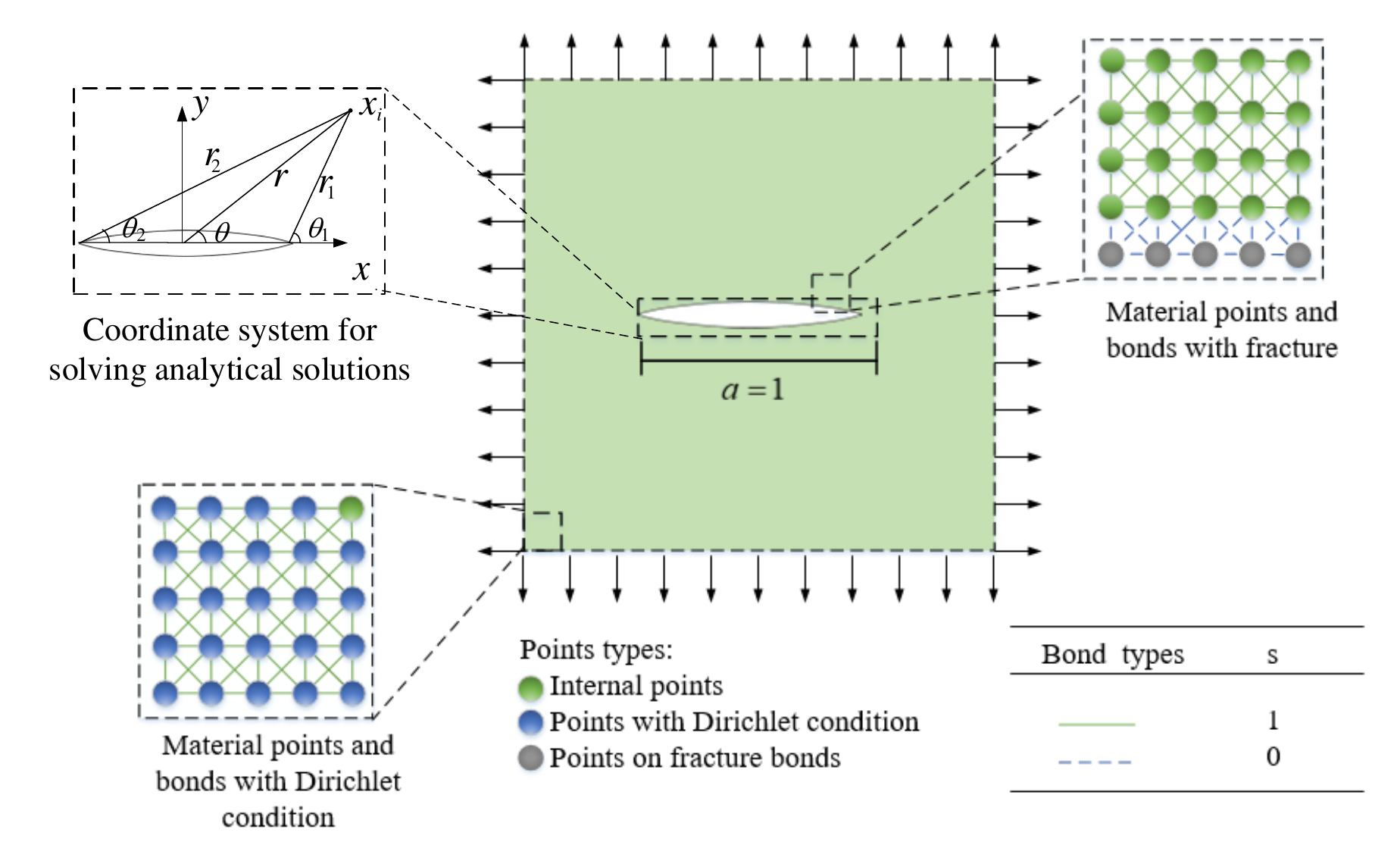}  
		\caption{Geometry of a plate with pre-existing crack.} 
		\label{Fig:7}
	\end{figure}

The previous subsection established the ability of BBPD-GK to handle continuous problems and ensured that any special numerical schemes do not cause AC convergence. In this subsection, we investigate the accuracy of BBPD-GK on the discontinuous problem on a plate with pre-existing crack.
Here we limit the region to $\Omega=[0,2]\times[0,2]$ and set analytical solutions $u_a$ as Dirichlet boundary conditions around it, as shown in Fig. \ref{Fig:7}.

 The crack length of the plane is chosen as $a=1$, and we select the material properties as Bulk modulus $\kappa=4\times 10^4$, Possion $\nu=1/4$. The displacement field $\mathbf{u}(\mathbf{x}_i)=(u(\mathbf{x}_i),v(\mathbf{x}_i))$ for this problem can be computed as
 \begin{equation}
      \begin{aligned}
      & 2Gu(\mathbf{x}_i,t)=(1-2\nu)  \sqrt{r_1 r_2} \cos \left(\frac{\theta_1+\theta_2}{2}\right)-\frac{ r^2}{\sqrt{r_1 r_2}} \sin \theta \sin \left(\theta-\frac{1}{2} \theta_1-\frac{1}{2} \theta_2\right), \\
& 2Gv(\mathbf{x}_i,t)=(2-2\nu)   \sqrt{r_1 r_2} \sin \left(\frac{\theta_1+\theta_2}{2}\right)-\frac{ r^2}{\sqrt{r_1 r_2}} \sin \theta \cos \left(\theta-\frac{1}{2} \theta_1-\frac{1}{2} \theta_2\right),   
     \end{aligned}
 \end{equation}
where $G$ is the shear modulus. The coordinates $r$, $r_1$, $r_2$, $\theta_1$  and $\theta_2$ are different for various material points $\mathbf{x}_i$, as shown in Fig. \ref{Fig:7}. 
 
\begin{figure}[ht]
	\centering    	
	\subfigure[]
	{
		\begin{minipage}{6cm}
			\centering          
			\includegraphics[scale=0.28]{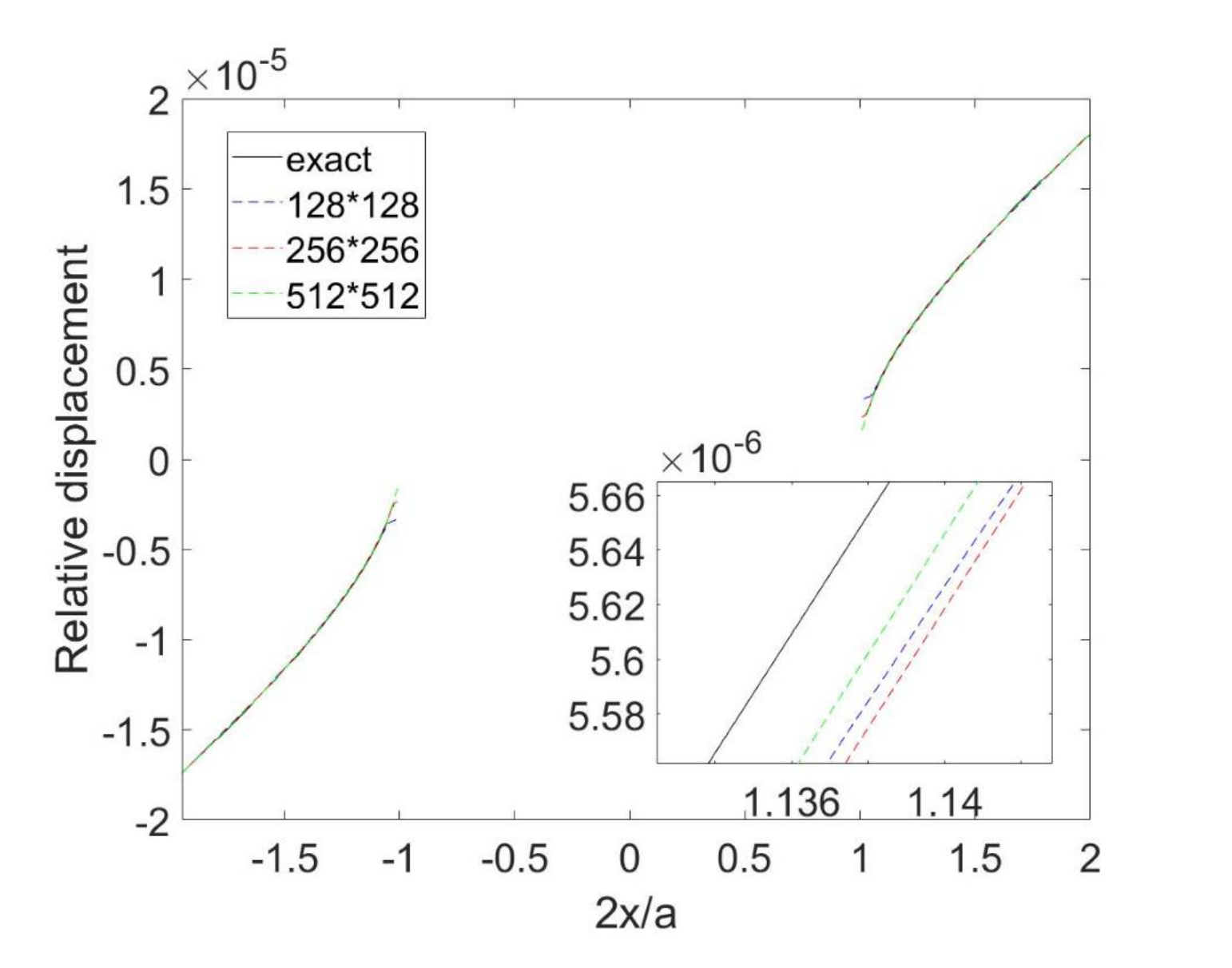}  
		\end{minipage}
	}	
	\subfigure[]
	{
		\begin{minipage}{6cm}
			\centering      
			\includegraphics[scale=0.28]{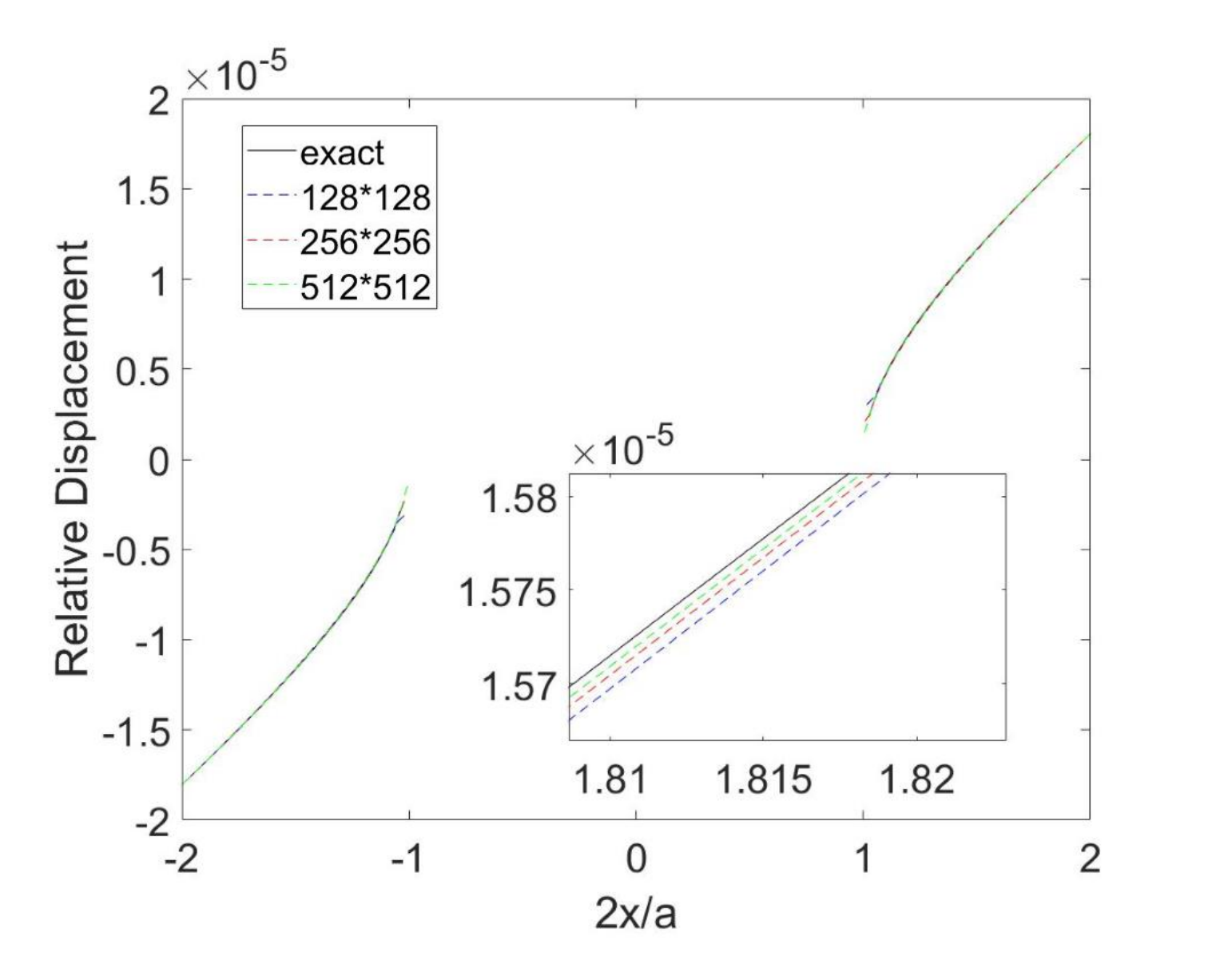}   
		\end{minipage}
	}	
	\caption{ Horizontal displacement along $y = 0$ axis computed by: (a) BBPD; (b) BBPD-GK.
} 
	\label{Fig:9}
\end{figure}

  The linear BBPD-GK and BB-PD with LAM method is used to simulate this model, where the spatial discretization is chosen as meshfree method. Keeping a fixed ratio of $\delta /\Delta x=3.5$ for BB-PD and $\sigma=\Delta x$ for BBPD-GK, we run this crack problem with $32^2, 64^2$ and $128^2$ points to verify the convergence. The numerical result obtained by two models when $y=0$ are shown in Figure \ref{Fig:9}. Here all displacements are scaled by $|u|_{L_{\infty}}$.

As expected, given results for the manufactured solution can be compatiable to the analysis solution when the grid size $\Delta x$ is changed from $1/32$ to $1/128$.
This verified the convergence of BBPD-GK in discontinuous problems. Additional, relative to BB-PD, BBPD-GK exhibits a higher degree of conformity with the displacement curve of the exact solution. Consequently, in scenarios characterized by discontinuous problems, BBPD-GK demonstrates its capacity to effectively mitigate computational errors arising from the intersection of neighborhood and material point volumes through direct discretization.
 
\subsection{Kalthoff-Winkler experiment}

\begin{figure}[ht]
		\centering            
		\includegraphics[scale=0.65]{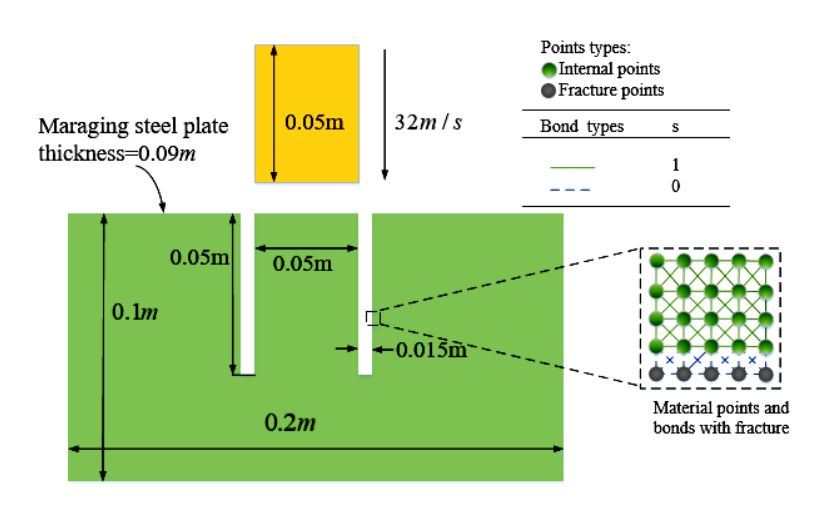}  
		\caption{Geometry of Kalthoff-Winkler experiment and its discretization.} 
		\label{Fig:12}
	\end{figure}

Finally, we consider a dynamic fracture problem. Here, we use an experiment called the Kalthoff-Winkler experiment to test, which involves using a cylinder to impact a block with a preset crack, as shown in Fig. \ref{Fig:12}.

The problem description is as follows: a block of length $L=0.2 \mathrm{~m}$, width $W=0.1 \mathrm{~m}$, and thickness $h=0.009 \mathrm{~m}$ with two thin notches is subjected to the incomplete horizons and impactor. The notch in this model has a width of $h_0=0.0015 \mathrm{~m}$, a length of $a_0=0.05 \mathrm{~m}$, and distance between notches $d$ is $0.05 \mathrm{~m}$. The impactor has a diameter of $D=0.05 \mathrm{~m}$ and a height of $H=0.05 \mathrm{~m}$.  The material properties are $E=191Gpa$ (elastic modulus), $v=1/4$ (Poisson's ratio) and $\rho=8000\mathrm{~kg}/\mathrm{m}^3$ (density), $\delta=1.5\times 10^{-4} \mathrm{~m}$ (horizon size), $s_0=0.01$ (critucal bond stretch). Initial velocity $v_0=32 \mathrm{~m}/\mathrm{s}$ is imposed to ensure crack propagation. The critical stretch for bond breaking is $s_0=0.01$. This simulation is conducted by the VV algorithm with the time step size $\Delta t=8.9 \times 10^{-8}s$.  For these properties, we obtain a bond-breaking criterion of $s_0=0.01$ following the \eqref{fra:s1}, and $s_c$ following the \eqref{sc}.

To verify nonlinear BBPD-GK's ability to reproduce crack propagation speed and branching location, we compare the experimental results of BBPD-GK and traditional BB-PD. All examples are obtained by meshfree discretizations and the velocity-Velet method. 
Here, the total number of material points is chosen as $201\times101\times 9$, and  $\Delta t=8.7\times 10^{-8}$ is selected to match the CFL condition imposed by the crack speed.

\begin{figure}[ht]
		\centering            
		\includegraphics[scale=0.35]{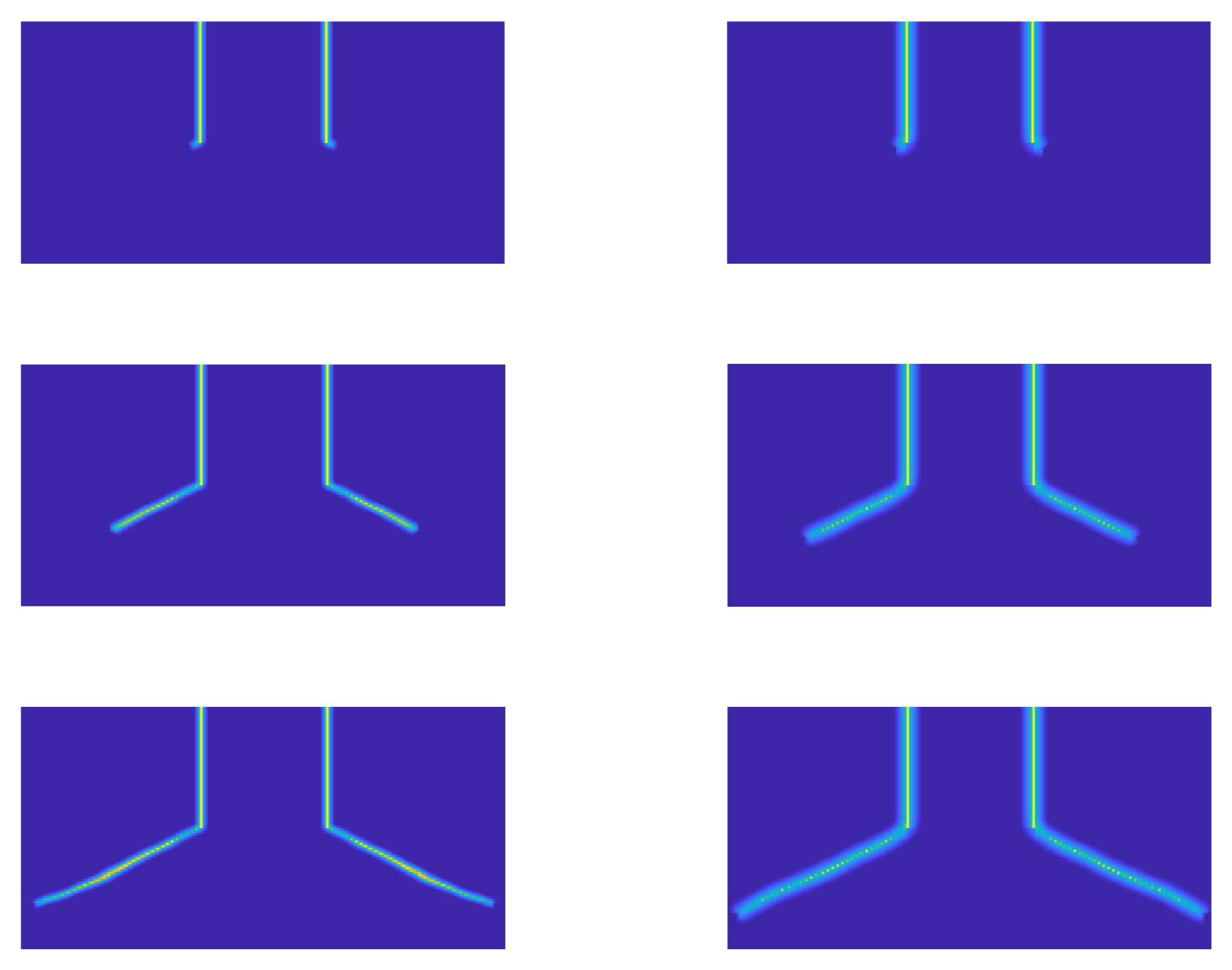}  
		\caption{. Crack evolution for Kalthoff-Winkler case after 350, 650, and 950 time steps obtained by the BB-PD (left) and BBPD-GK (right).} 
		\label{Fig:13}
	\end{figure}

Fig. \ref{Fig:13} shows the crack propagation when the number of time steps is chosen as $350$, $650$, and $950$, where the horizon size $\delta$ of BB-PD is chosen as $3.015\Delta x$, and the parameter $\sigma$ of BBPD-GK is chosen as $\Delta x$. Here, the blue area indicates that there is no fracture between the material points, i.e., $s=1 $; The yellow area indicates that the bonds between the material points have broken, i.e., $s=0 $. In other regions, material points undergo deformation but not complete fracture, with a value of $s$ ranging from 0 to 1, which we call partial bond breakage. 

 Ideally, we hope to recover results using our model that are comparable to traditional BB-PD on uniform discretizations without the need to introduce volume correction. By comparing the corresponding fracture patterns in Fig. \ref{Fig:13}, we can obverse that branching happens when the number of time steps is chosen as $650$, and both models predict approximately the same $65$ degree between primary crack branches, which verifies that the proposed BBPD-GK can accurately capture the fracture behavior of solid materials.
 \begin{figure}[ht]
	\centering    	
	\subfigure[]
	{
		\begin{minipage}{5cm}
			\centering      
			\includegraphics[scale=0.23]{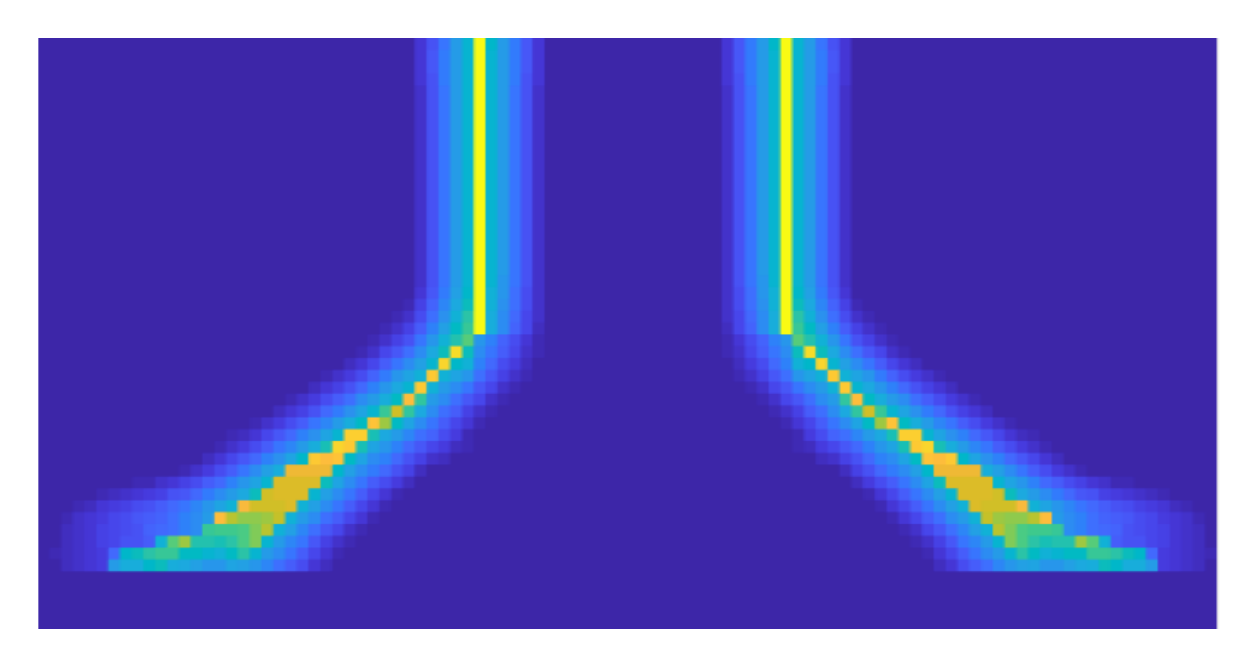}   
		\end{minipage}
	}	
 	\subfigure[]
	{
		\begin{minipage}{5cm}
			\centering          
			\includegraphics[scale=0.3]{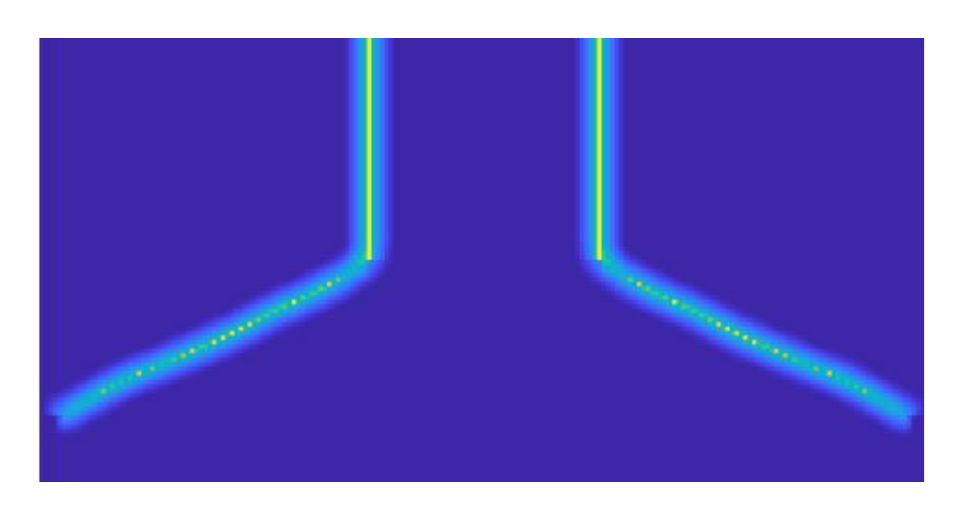}  
		\end{minipage}
	}	
	\subfigure[]
	{
		\begin{minipage}{5cm}
			\centering      
			\includegraphics[scale=0.3]{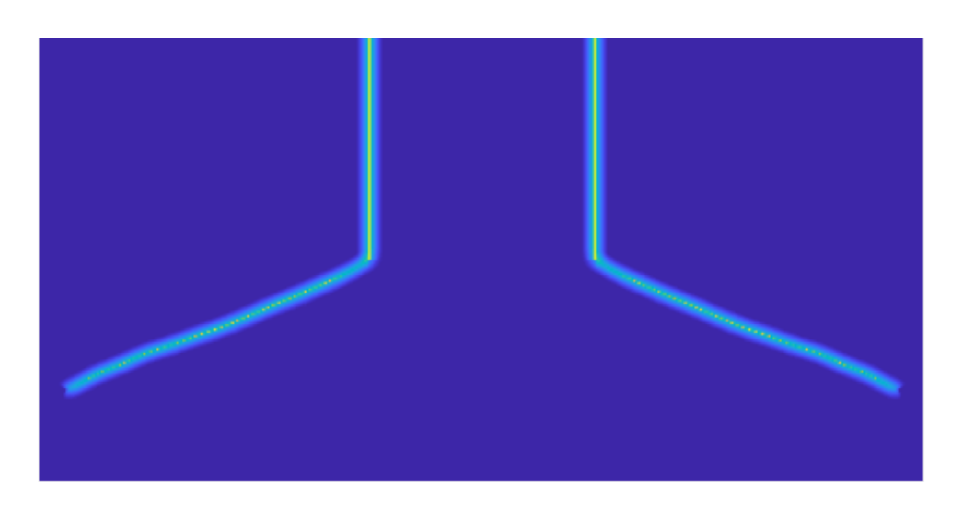}   
		\end{minipage}
	}	
	\caption{Crack evolution for Kalthoff-Winkler case with different resolution: (a)$N=101\times51$; (b)$N=201\times101$; (c)$N=401\times201$.
} 
	\label{Fig:14}
\end{figure}

 We illustrate convergence of the fully resolved dynamics in  Fig. \ref{Fig:14} by fix ing $\sigma=\Delta x$ and $\chi_a^2(2)=36$ and increase the number of material points.  As the number of material points increases, the observed crack angles are $51$ degrees in Fig. \ref{Fig:14}(a), 65 degrees in Fig. \ref{Fig:14}(b), and 66 degrees in Fig. \ref{Fig:14}(c). This means that although we cannot obtain accurate convergence analysis in discontinuous problems, we can observe a converged fragment shape that reproduces the experimentally observed crack angle at the pre-notch tip. 


 \begin{figure}[ht]
	\centering    	
 	\subfigure[]
	{
		\begin{minipage}{7cm}
			\centering          
			\includegraphics[scale=0.3]{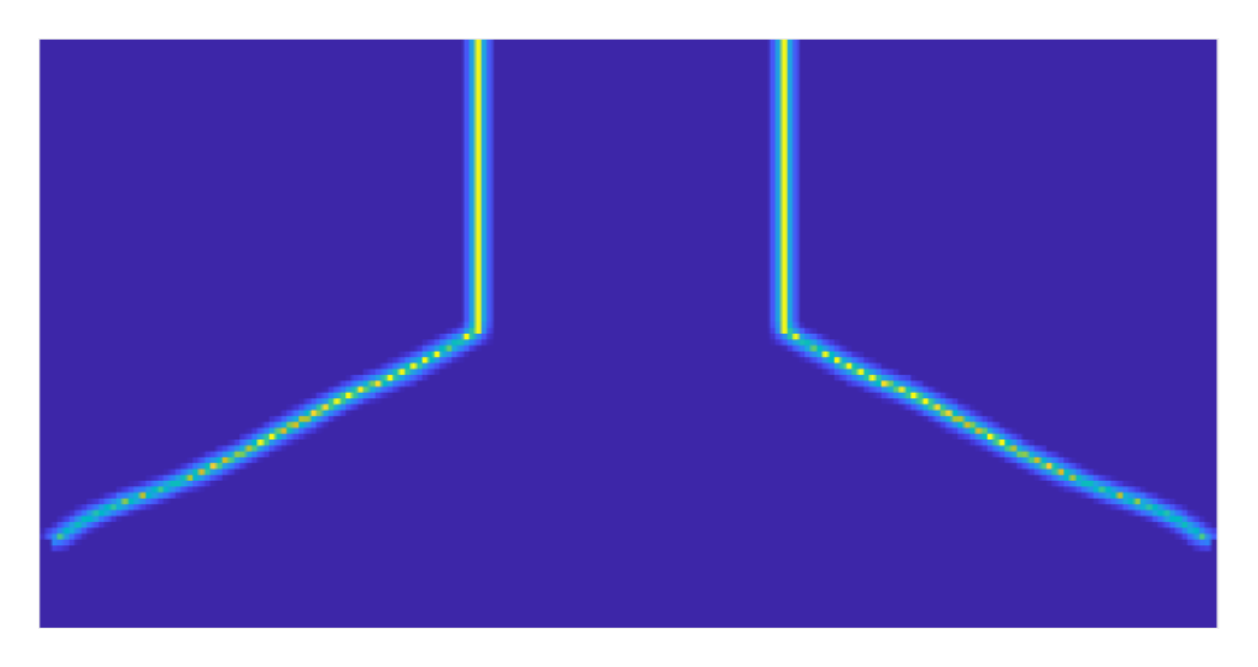}  
		\end{minipage}
	}	
	\subfigure[]
	{
		\begin{minipage}{7cm}
			\centering      
			\includegraphics[scale=0.3]{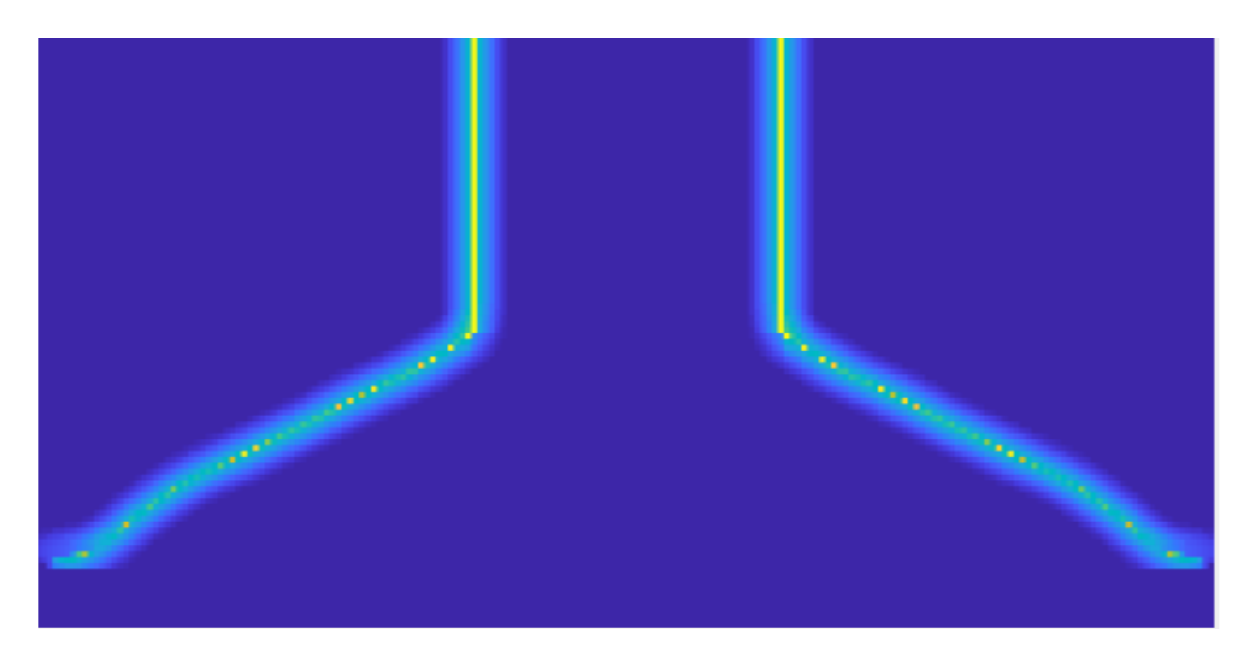}   
		\end{minipage}
	}	
	\caption{Crack evolution for Kalthoff-Winkler case: (a)By linear BB-PD; (b)By linear BBPD-GK; 
} 
	\label{Fig:16}
\end{figure}
Based on the previous peridynamic presentation, under the assumption of small deformation, linear models can approximate the effects of nonlinear models. Thus, we test the perform of the linear BB-PD and BBPD-GK with the material points is chosen as $201\times101\times 9$. As shown in Fig. \ref{Fig:16}, both linear BBPD-GK and BB-PD obtain a crack angle consistent with the nonlinear model, which proves this conclusion.

\begin{figure}[ht]
	\centering    	
	\subfigure[]
	{
		\begin{minipage}{5cm}
			\centering      
			\includegraphics[scale=0.23]{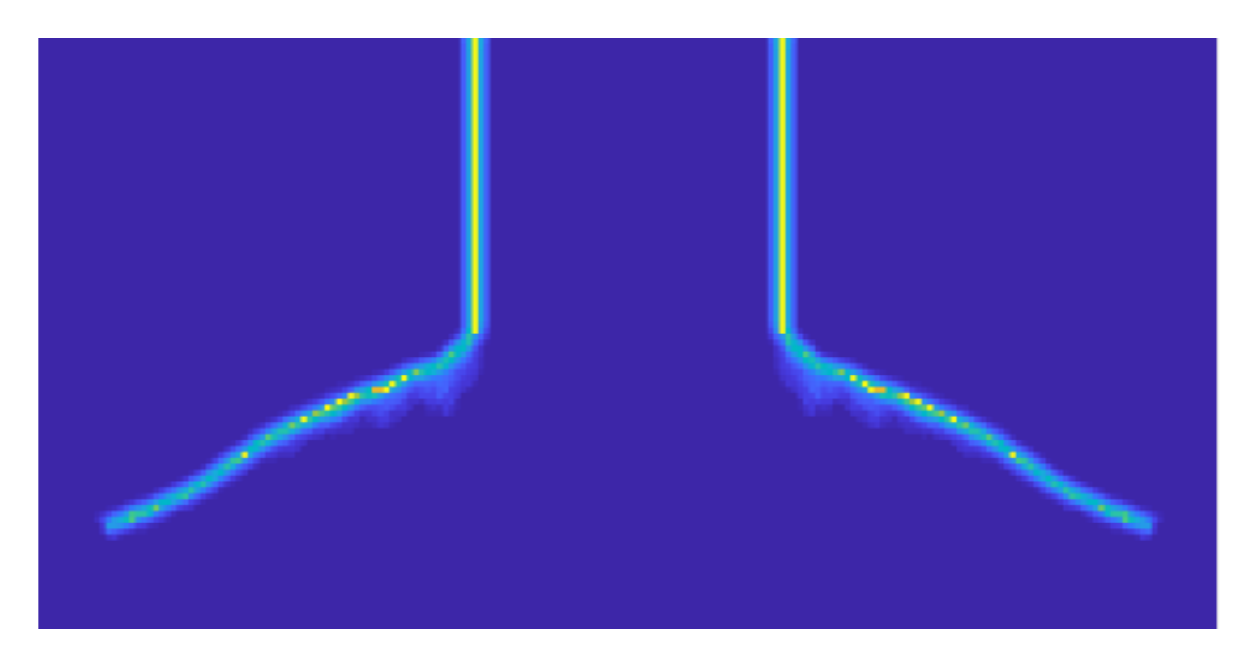}   
		\end{minipage}
	}	
 	\subfigure[]
	{
		\begin{minipage}{5cm}
			\centering          
			\includegraphics[scale=0.3]{NOLIN201NEW.pdf}  
		\end{minipage}
	}	
	\subfigure[]
	{
		\begin{minipage}{5cm}
			\centering      
			\includegraphics[scale=0.3]{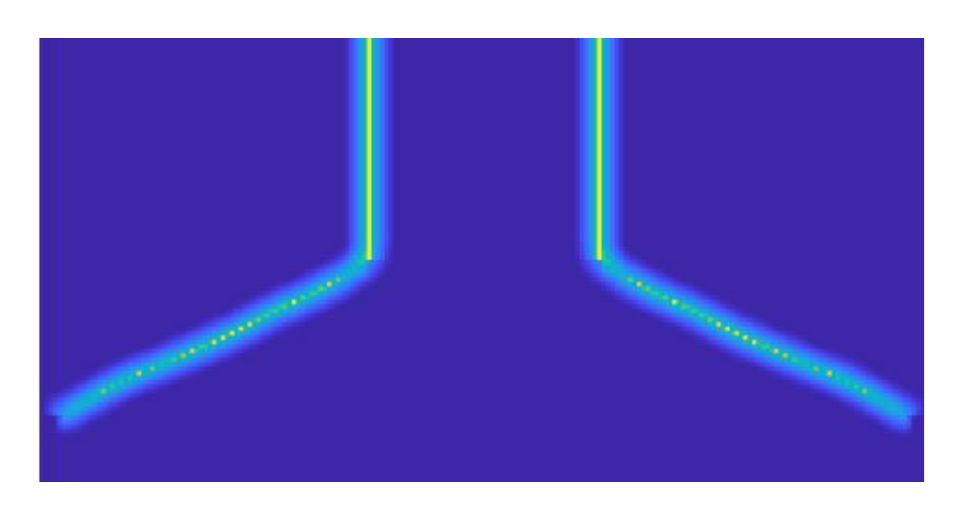}   
		\end{minipage}
	}	
	\caption{Crack evolution for Kalthoff-Winkler case with different $\sigma$: (a)$\sigma=\Delta x/2$; (b) $\sigma=\Delta x$; (c) $\sigma=2\Delta x$.
} 
	\label{Fig:17}
\end{figure}

\begin{figure}[ht]
	\centering  
  	\subfigure[]
	{
		\begin{minipage}{6cm}
			\centering      
			\includegraphics[scale=0.3]{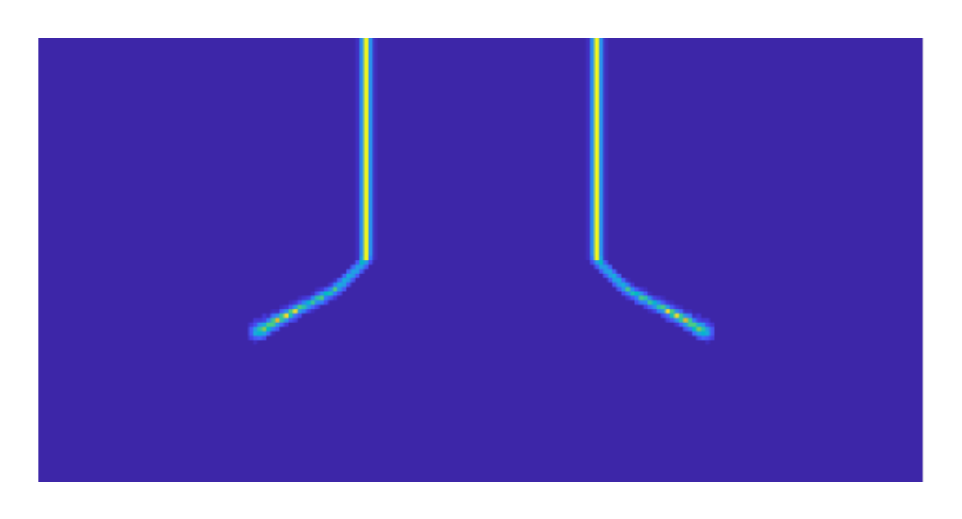}   
		\end{minipage}
	}
 	\subfigure[]
	{
		\begin{minipage}{6cm}
			\centering      
			\includegraphics[scale=0.3]{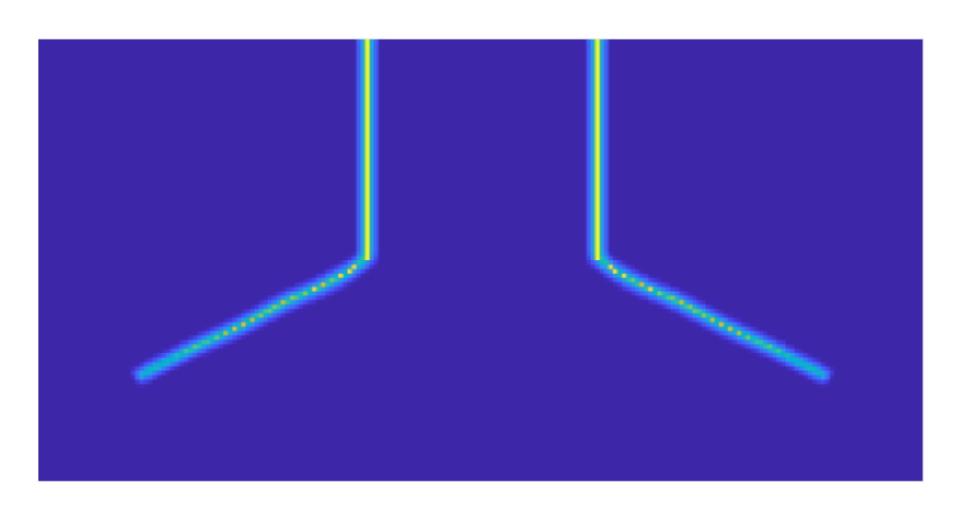}   
		\end{minipage}
	}
 	\subfigure[]
	{
		\begin{minipage}{6cm}
			\centering      
			\includegraphics[scale=0.3]{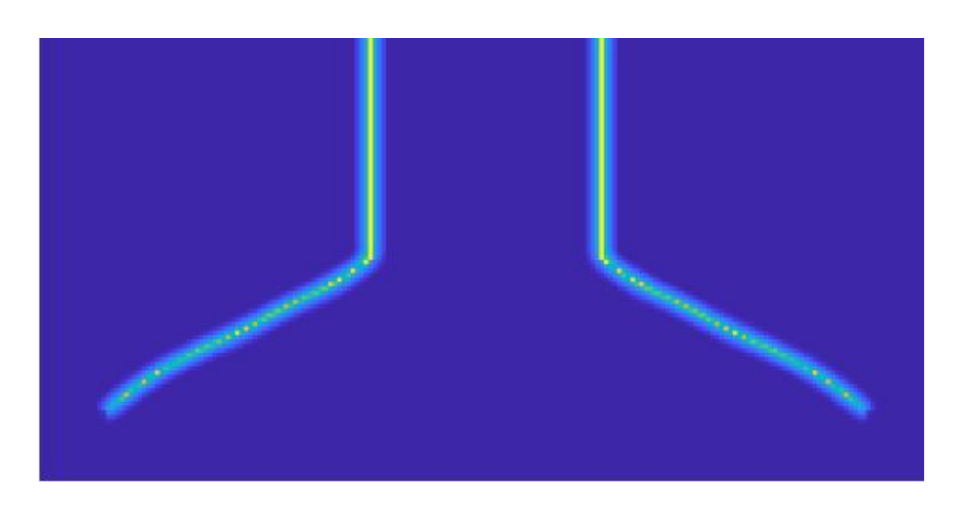}   
		\end{minipage}
	}
	\subfigure[]
	{
		\begin{minipage}{6cm}
			\centering      
			\includegraphics[scale=0.22]{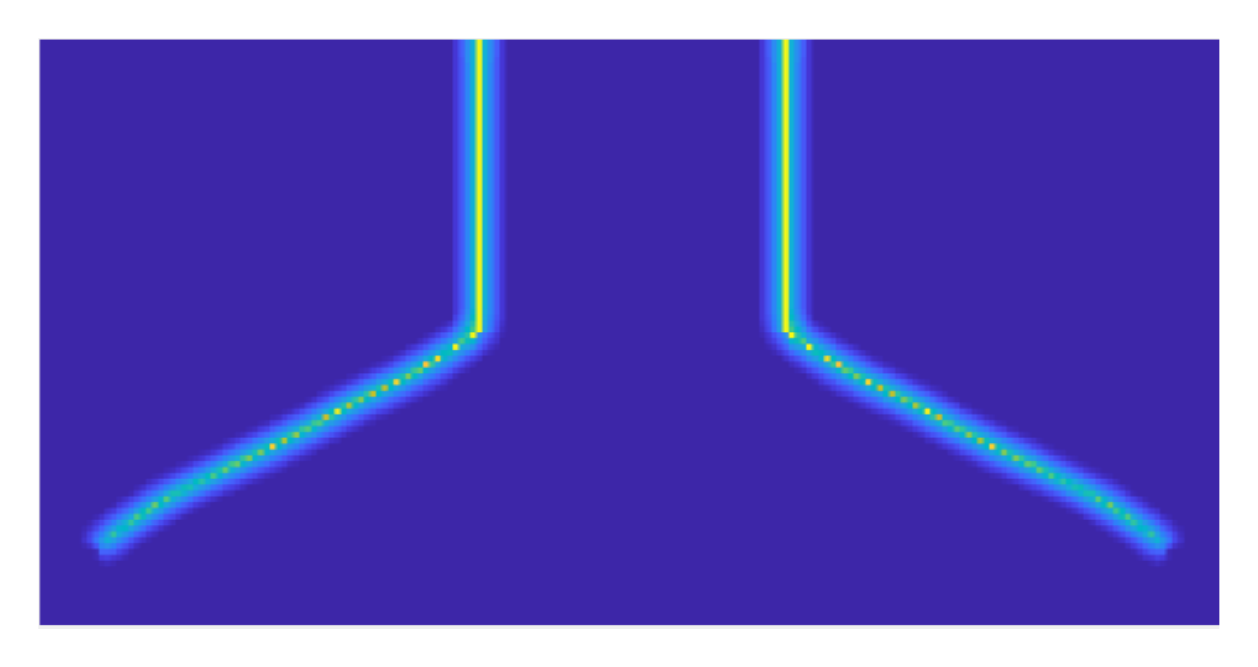}   
		\end{minipage}
	}	
 	\subfigure[]
	{
		\begin{minipage}{6cm}
			\centering          
			\includegraphics[scale=0.3]{NOLIN201NEW.pdf}  
		\end{minipage}
	}	
	\subfigure[]
	{
		\begin{minipage}{6cm}
			\centering      
			\includegraphics[scale=0.22]{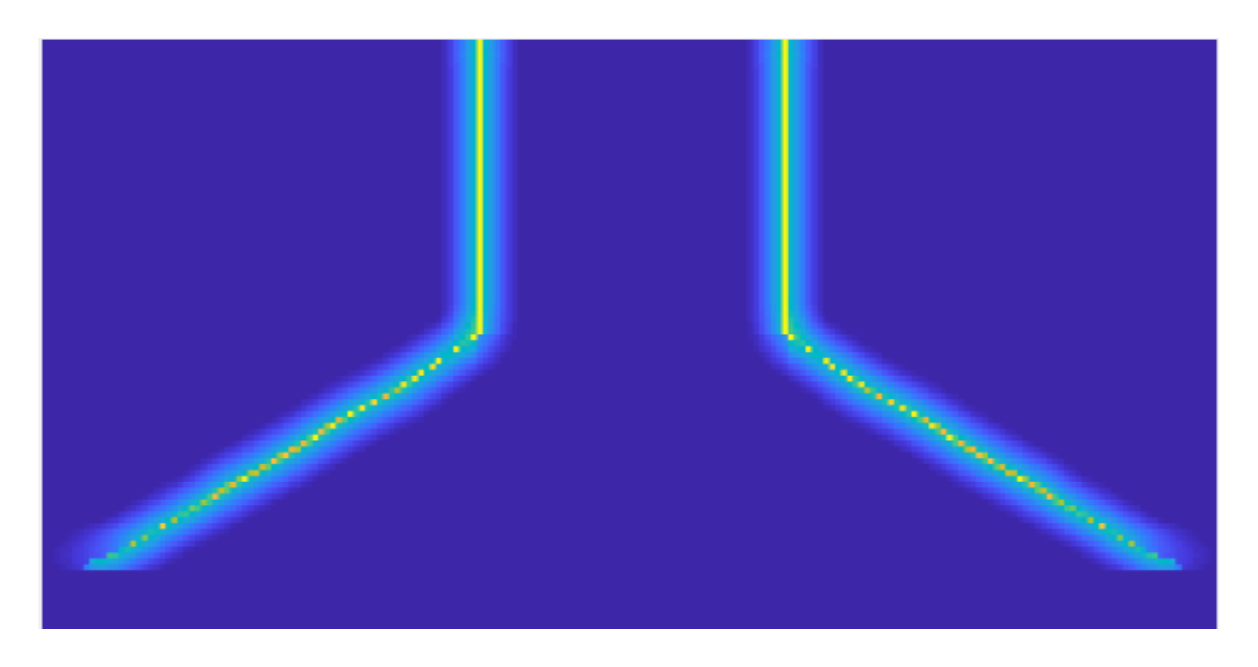}   
		\end{minipage}
	}	
	\caption{Crack evolution for Kalthoff-Winkler case with different $\chi_a^2(2)$: (a)$\chi_a^2(2)=4$; (b) $\chi_a^2(2)=9$; (c) $\chi_a^2(2)=16$; (d)$\chi_a^2(2)=25$; (e) $\chi_a^2(2)=36$; (f) $\chi_a^2(2)=49$.
} 
	\label{Fig:18}
\end{figure}

To better understand the crack mechanism presented in this problem, the effects of increasing the $\sigma$ and $\chi_a^2(2)$ in our BBPD-GK model are examined. Here we illustrate the peridynamics values by fixing $\chi_a^2(2)=36$, material points $N=201\times101\times 9$ and changing the $\sigma$ from $\Delta x/2$ to $2\Delta x$, as shown in Fig. \ref{Fig:17}. Based on the above-mentioned description of BBPD-GK for smooth problems, when $\sigma$ is selected as $\Delta x$, the simulation accuracy is significantly higher than when $\sigma$ is selected as $\Delta x/2$, and the simulation results of cracks confirm this. Meanwhile, another noteworthy observation is that as $\sigma$ is selected to a larger value, such as $2\Delta x$, the accuracy of the simulation only improves slightly, making $\sigma=\Delta x$ a more reasonable choice. Finally, in Fig. \ref{Fig:18}, we present crack profiles obtained with $\sigma=\Delta x$ but for varying truncation of the influence region. It can be observed that, compared to the crack simulation when $\chi_a^2(2)=4, 9, 16, 25$, the simulations with $\chi_a^2(2)=36$ and $\chi_a^2(2)=49$ in the BBPD-GK model produce crack evolution consistent with standard experimental results, providing evidence for the selection of $\chi_a^2(2)=36$.


By comparing BB-PD and BBPD-GK, we found that our proposed model exhibits asymmetric compatibility with the corresponding linear elastic local solution in continuous problems without requiring additional corrections. Moreover, it achieves crack simulation results consistent with standard experiments in crack propagation problems. The experiment also verified the rationality of selecting the parameters $\sigma = \Delta x$ and $\chi_a^2(2) = 36$ for the proposed model.
\section{Conclusions}
The primary contribution of this paper is the introduction of a novel bond-based peridynamics model leveraging Gaussian functions. This model addresses the convergence and volume correction challenges inherent in traditional peridynamics models. Traditional bond-based peridynamics models has been shown to exhibit a layer of asymmetric compatibility to the corresponding linear elastic local solution, and have significant difficulties in achieving accurate domain volume calculations and necessitating numerous correction methods. By employing a kernel function that operates over an unbounded region, our proposed model achieves asymptotic compatibility through direct meshfree discretization. This approach enables precise volume calculations for the neighborhood of material points without the need for additional correction methods.
﻿
Through energy equivalence, we establish the relationship between the proposed Gaussian Kernel based peridynamic (BBPD-GK) model and traditional peridynamic (PD) models, offering new fracture and linear models. The proposed BBPD-GK model demonstrates superior convergence and accuracy in test problems with known solutions. Furthermore, it maintains its convergence and accuracy in crack propagation problems, as evidenced by a series of discontinuous problem scenarios.
   \section*{Acknowledgments}
   The first author (Hao Tian) was supported by the Fundamental Research Funds for the Central Universities (Nos. 202042008 and 202264006) and the National Natural Science Foundation of China (No. 11801533).
 \bibliographystyle{elsarticle-num}
\bibliography{refer}
\end{document}